\documentclass[]{aa}  

\usepackage{natbib}
\bibpunct{(}{)}{;}{a}{}{,}
\usepackage{graphicx}
\usepackage{revsymb}	
\usepackage{amsmath}	
\usepackage[usenames]{color}	
\usepackage{epstopdf}	
\usepackage{txfonts}
\usepackage{amssymb}
\usepackage{color}
\usepackage[justification=centering]{caption}
\usepackage{geometry} 
\usepackage[parfill]{parskip} 
\usepackage{amssymb}
\usepackage{slantsc}
\usepackage{array}
\usepackage{url}
\usepackage{amsfonts}
\usepackage[colorlinks=true,linkcolor=black, urlcolor=black, citecolor=black]{hyperref}
\usepackage{hyperref} 
\usepackage{sidecap}
\usepackage{nicefrac}
\usepackage{subfigure}
\usepackage{epsfig}
\usepackage{xcolor}
\colorlet{bleu}{blue!90!darkgray}
\colorlet{rouge}{red!70!darkgray}
\colorlet{vert}{green!60!black}
\colorlet{gris}{black!50}

\begin{document}
\title{Using seismic inversions to obtain an internal mixing processes indicator for main-sequence solar-like stars}
\author{G. Buldgen\inst{1}\and D. R. Reese\inst{2}\and M. A. Dupret\inst{1}}
\institute{Institut d’Astrophysique et Géophysique de l’Université de Liège, Allée du 6 août 17, 4000 Liège, Belgium \and School of Physics and Astronomy, University of
Birmingham, Edgbaston, Birmingham, B15 2TT, UK.}
\date{April, 2015}
\abstract{Determining accurate and precise stellar ages is a major problem in astrophysics. These determinations are either obtained through empirical relations or model-dependent approaches. Currently, seismic modelling is one of the best ways of building accurate stellar models, and therefore providing accurate ages. However, current methods are affected by simplifying assumptions concerning stellar mixing processes. In this context, providing new structural indicators which are less model-dependent and more sensitive to mixing processes is crucial.}
{We wish to build a new indicator for core conditions (i.e. mixing processes and evolutionary stage) on the main sequence. This indicator \textbf{$t_{u}$} should be more sensitive to structural differences and applicable to older stars than the indicator $t$ presented in a previous paper. We also wish to analyse the importance of the number and type of modes for the inversion, as well as the impact of various constraints and levels of accuracy in the forward modelling process that is used to obtain reference models for the inversion.}
{First, we present a method to obtain new structural kernels in the context of asteroseismology. We then use these new kernels to build a new indicator of central conditions in stars, denoted $t_{u}$, and test it for various effects including atomic diffusion, various initial helium abundances and various metallicities, following the seismic inversion method presented in our previous paper. We then study its accuracy for 7 different pulsation spectra including those of $16$CygA and $16$CygB and analyse how it depends on the reference model by using different constraints and levels of accuracy for its selection}
{We observe that the inversion of the \textbf{new} indicator $t_{u}$ using the SOLA method provides a good diagnostic for additional mixing processes in central regions of stars. Its sensitivity allows us to test for diffusive processes and chemical composition mismatch. We also observe that using modes of degree $3$ can improve the accuracy of the results, as well as using modes of low radial order. Moreover, we note that individual frequency combinations should be considered to optimize the accuracy of the results.}{}
\keywords{Stars: interiors -- Stars: oscillations -- Stars: fundamental parameters -- Asteroseismology}
\maketitle
\section{Introduction}
Determining stellar ages accurately is crucial when studying stellar evolution, determining properties of exoplanetary systems or characterising stellar populations in the galaxy. However, the absence of a direct observational method to measure this quantity makes such determinations rather complicated. Age is usually related empirically to the evolutionary stage or determined through model-dependent techniques like forward asteroseismic modelling of stars. This model-dependence is problematic since if a physical process is not taken into account during the modelling, we will introduce a bias in the age determinations, as well as in the determination of other fundamental characteristics like the mass or the radius (see for example \citet{Eggenberger} for the impact of rotation on asteroseismic properties, \citet{Miglio} for a discussion in the context of ensemble asteroseismology and \citet{Brown} for a comprehensive study of the relation between seismic constraints and stellar model parameters). It is also clear that asteroseismology probes the evolutionary stage of stars and not directly the age. In other words, we are able to analyse the stellar physical conditions but relating these properties to an age will, ultimately, always be dependent on the assumptions made during the building of the evolutionary sequence of the model. A general review of the impact of the hypotheses of stellar modelling and of asteroseismic constraints on the determination of stellar ages is presented in \citet{Lebretona} and \citet{Lebretonb}. 
\\
\\
In that sense, the question of additional mixing processes in the context of seismic modelling is central \citep{Dupret} and can only be solved by using less model-dependent seismic analysis techniques and new generations of stellar models. These new seismic methods should be able to provide relevant constraints on the physical conditions in the central regions and help with the inclusion of additional mixing in the models. In this context, seismic inversion techniques are an interesting way to relate structural differences to frequency differences and therefore offer a new insight on the physical conditions inside observed stars. From the observational point of view, the high quality of the Kepler and CoRoT data but also the selection of the Plato mission \citep{Rauer} allows us to expect enough observational data to carry out inversions of global characteristics. In the context of helioseismology, structural inversion techniques have already lead to noteworthy successes. They have provided strong constraints on solar atomic diffusion \citep[see][]{Basudiff}, thus confirming the work of \citet{Elsworth}. However, their application in asteroseismology is still limited. Inversions for rotation profiles have been carried out \citep[see for example][for an application to Kepler subgiants]{Deheuvels}, but as far as structural inversions are concerned, one can use either non-linear inversion techniques \citep[see][for an example of the differential response technique]{RoxTheory, RoxApplication}, or linear inversion techniques applied to integrated quantities as is done in \citet{Reese} and \citet{Buldgen}. 
\\
\\
In our previous paper \citep[see][]{Buldgen}, we extended mean density inversions based on the SOLA technique \citep{Reese} to inversions of the acoustic radius of the star and an indicator of core conditions, denoted $t$. We also set the basis of a general approach to determine custom-made global characteristics for an observed star. We showed that applying the SOLA inversion technique \citep{Pijpers} to a carefully selected reference model, obtained via the forward modelling technique, could lead to very accurate results. However, it was then clear that the first age indicator was limited to rather young stars and that other indicators should be developed. Moreover, the model-dependence of these techniques should be carefully studied and there is a need to define a more extended theoretical background for these methods. The influence of the number but also the type of modes used for a specific inversion should be investigated. In the end, one should be able to define whether the inversion should be carried out or not, knowing the number of observed frequencies and the quality of the reference model according to its selection criteria. 
\\
\\
In this study, we wish to offer an answer to these questions and provide a new indicator for the mixing processes and the evolutionary stage of an observed star. We structure our study as follows: section \ref{secnewkernels} introduces a technique to obtain equations for new structural kernels in the context of asteroseismology and applies it to the $(u,\Gamma_{1})$ and the $(u,Y)$ kernels, where $u$ is the squared isothermal sound speed, $\Gamma_{1}$ the adiabatic gradient and $Y$ the current helium abundance profile. Section \ref{secnewindic} introduces a new indicator of mixing processes and evolutionary stages, which is not restricted to young stars, as was the case for the indicator presented in \citet{Buldgen}. Having introduced this new indicator, we test its accuracy using different physical effects such as including atomic diffusion processes with high velocities (up to $2.0$ times the solar microscopic diffusion velocities) in the target, changing the helium abundance, changing the metallicity and changing the solar mixture of heavy elements. Section \ref{secmodestudy} analyses the impact of the type and number of modes on the inversion results whereas Sect. \ref{secmodelstudy} studies how the accuracy depends on the reference model. We also tested our method on targets similar to the binary system $16$CygA and $16$CygB using the same modes as those observed in \citet{Verma} to show that our method is indeed applicable to current observational data. Section \ref{secconclusion} summarizes our results and presents some prospects on future research for global quantities that could be obtained with the SOLA inversion technique.
\section{$(u,\Gamma_{1})$ and $(u,Y)$ structural kernels}\label{secnewkernels}
\subsection{Integral equations for structural couples in the asteroseismic context}\label{secscalingtest}
\begin{figure*}[t]
	\centering
		\includegraphics[width=17.5cm]{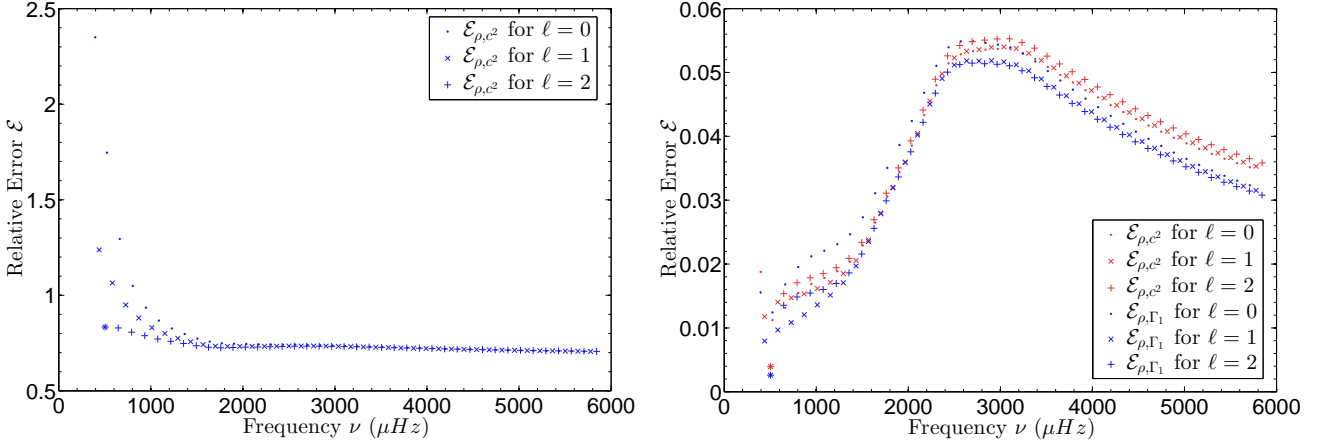}
	\caption{Verification of Eq. \ref{eqrhoc2couple} for a set of $120$ modes with the $\cdot$ being the $40$ radial modes, the $\times$ being the $40$ dipolar modes and the $+$ being the $40$ quadrupolar modes. The left plot illustrates this verification for the $(\rho,c^{2})$ couple when no scaling is applied to $c^{2}$ whereas the right plot illustrates the same verification for the $(\rho,c^{2})$ couple when scaling is applied to $c^{2}$, as well as the $(\rho,\Gamma_{1})$ couple where no scaling is needed.}
		\label{figverifscaling}
\end{figure*}

It has been demonstrated in \citet{Gough} that one could deduce from the variational principle a linear integral relation between the perturbations of frequencies and the perturbation of structural variables. This equation is obtained by assuming the adiabatic approximation and spherical symmetry, and neglecting surface integral terms. It is only valid if the stellar models are sufficiently close to each other. If one is working with the structural pair $(c^{2},\rho)$, where $c^{2}$ is the adiabatic squared sound velocity and $\rho$ the density, this relation takes on the following form:
\begin{align}
\frac{\delta \nu_{i}}{\nu_{i}}=\int_{0}^{1}K^{i}_{\rho,c^{2}}\frac{\delta \rho}{\rho}dx + \int_{0}^{1}K^{i}_{c^{2},\rho}\frac{\delta c^{2}}{c^{2}}dx + \frac{\mathcal{G}(\nu)}{Q_{i}}+\mathcal{O}(2), \label{eqrhoc2couple}
\end{align}
where $x=\frac{r}{R}$ with $R$ the stellar radius, and where the classical definition of the relative differences between the target and model for any structural quantity $s$ has been used:
\begin{align}
\frac{\delta s(x)}{s}=\frac{s_{\mathrm{obs}}(x)-s_{\mathrm{ref}}(x)}{s_{\mathrm{ref}}}. \label{eqtargetrefstruct}
\end{align}
In what follows, we will always use the subscript or superscript ``$\mathrm{obs}$'' when referring to the observed star, ``$\mathrm{ref}$'' for the reference model variables in perturbation definitions and $\mathrm{inv}$ for inverted results. Other variables, such as the kernel functions, denoted without subscripts or superscripts, are of course related to the reference model and are known in practice. Finally, one should also note that the suffix $i$ is just a index to classify the modes. Moreover, since it is clear that some hypotheses are not suitable for surface regions, a supplementary function, $\mathcal{G}(\nu)$ was added to model these so-called surface effects. It is defined as a linear combination of Legendre polynomials, normalised by the factor $Q_{i}$, which is the mode inertia normalised by the inertia of a radial mode interpolated to the same frequency. We emphasize here that neither this normalisation coefficient nor the treatment of surface effects are uniquely defined and other techniques are also used \citep[see for example][]{Dziembowski,Daeppen,Basusurf}. 
\\
\\
The kernels of the couple $(\rho,\Gamma_{1})$ were already presented in \citet{Gough} who also mentionned the use of another method, defined in \citet{Masters} to modify Eq. \ref{eqrhoc2couple} and obtain such relations for the $(c^{2},\Gamma_{1})$ couple and also the $(N^{2},c^{2})$ couple. Other approaches to obtain new structural kernels were presented  \citep[see for example][for the application of the adjoint equations method to this problem]{Elliott, Kosovichev}. This latter approach has been used in helioseismology where it was assumed that the mass of the observed star is known to a sufficient level of accuracy to impose surface boundary conditions. In the context of asteroseismology, we cannot make this assumption. Nevertheless, the approach defined in \citet{Masters} allows us to find ordinary differential equations for a large number of supplementary structural kernels, without assuming a fixed mass \footnote{The method of adjoint equations previously described could also be used but would require an additional hypothesis to replace the missing boundary condition.}. 
\\
\\
Another question arises in the context of asteroseismology: what about the radius? We implicitely define our integral equation in non-dimensional variables but how do we relate the structural functions, for example $c^{2}_{\mathrm{obs}}(r)$ defined for the observed star and $c^{2}_{\mathrm{ref}}(r)$ defined for the reference model? What are the implications of defining all functions on the same domain in $x=\frac{r}{R_{ref}}$ varying from $0$ to $1$? It was shown by \citet{Basusca} that an implicit scaling was applied by the inversion in the asteroseismic context. The observed target is homologously rescaled to the radius of the reference model, while its mean density is preserved. This means that the oscillation frequencies are the same, but other quantities such as the adiabatic sound speed $c$, the squared isothermal sound speed $u=\frac{P}{\rho}$ will be rescaled. Therefore, when inverted, they are not related to the real target but to a scaled one.
\\
\\
This can be demonstrated with the following simple test. We can take two models a few time steps from one another on the same evolutionary sequence knowing that they should not be that different (here, we consider $1M_{\odot}$, main-sequence models). We then test the verification of Eq. \ref{eqrhoc2couple} by plotting the following relative difference:
\begin{align}
\mathcal{E}^{i}_{\rho,c^{2}}=\frac{\frac{\delta \nu_{i}}{\nu_{i}}-\mathcal{S}^{i}}{\frac{\delta \nu_{i}}{\nu_{i}}}, \label{eqverifvar}
\end{align}
with $\mathcal{S}^{i}$ defined as follows:
\begin{align}
\mathcal{S}^{i}=\int_{0}^{1}\left(K^{i}_{\rho,c^{2}}\frac{\delta \rho}{\rho} +K^{i}_{c^{2},\rho}\frac{\delta c^{2}}{c^{2}}\right)dx. \label{strucvar}
\end{align}
The results are plotted in blue in the left panel of Fig. \ref{figverifscaling}, where we can see that this equation is not satisfied. However, one could think that this inaccuracy is related to the neglected surface terms or to non-linear effects. Therefore we carry out the same test using the $(\rho,\Gamma_{1})$ kernels, plotted in red in the right panel of Fig. \ref{figverifscaling}. We see that for these kernels, the equation is satisfied. Moreover, when separating the contributions of each structural term, we see that the errors arise from the term related to $c^{2}$ in the first case. Using the scaled adiabatic sound speed, however, leads to the blue symbols in the right panel of Fig. \ref{figverifscaling} and we directly see that in this case, the integral equation is satisfied. This leads to the conclusion that inversion results based on integral equations are always related to the scaled target and not the target itself, as was concluded by \citet{Basusca}. We will see in Sect. \ref{secnewindic} that this has strong implications on the structural information given by inversion techniques.
\subsection{Differential equation for the $(u,\Gamma_{1})$ and the $(u,Y)$ kernels}
\begin{figure*}[t]
	\centering
		\includegraphics[width=17.5cm]{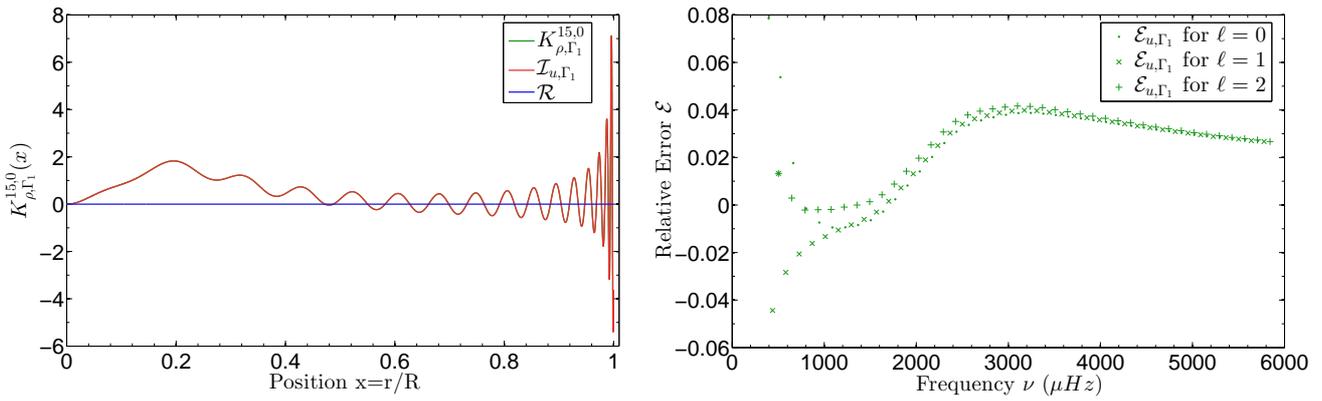}
	\caption{The left plot illustrates the verification of Eq. \ref{eqkernelugamma} for $n=15$, $\ell=0$ kernel noted $K^{15,0}$, where $\mathcal{I}_{u,\Gamma_{1}}$ is the right hand side of this equation and $\mathcal{R}$ is the residual. The right plot illustrates the same test as in Fig. \ref{figverifscaling} for the $(u,\Gamma_{1})$ kernels.}
		\label{figverifkerugam}
\end{figure*}

\begin{figure*}[t]
	\centering
		\includegraphics[width=17.5cm]{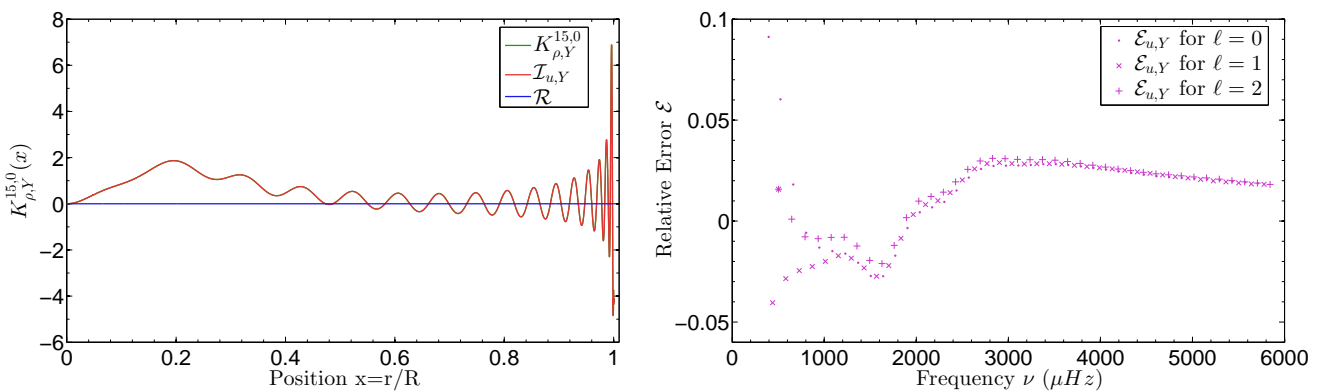}
	\caption{Same as Fig. \ref{figverifkerugam} for the $(u,Y)$ kernels.}
		\label{figverifkeruy}
\end{figure*}

\begin{figure*}[t]
	\centering
		\includegraphics[width=17.5cm]{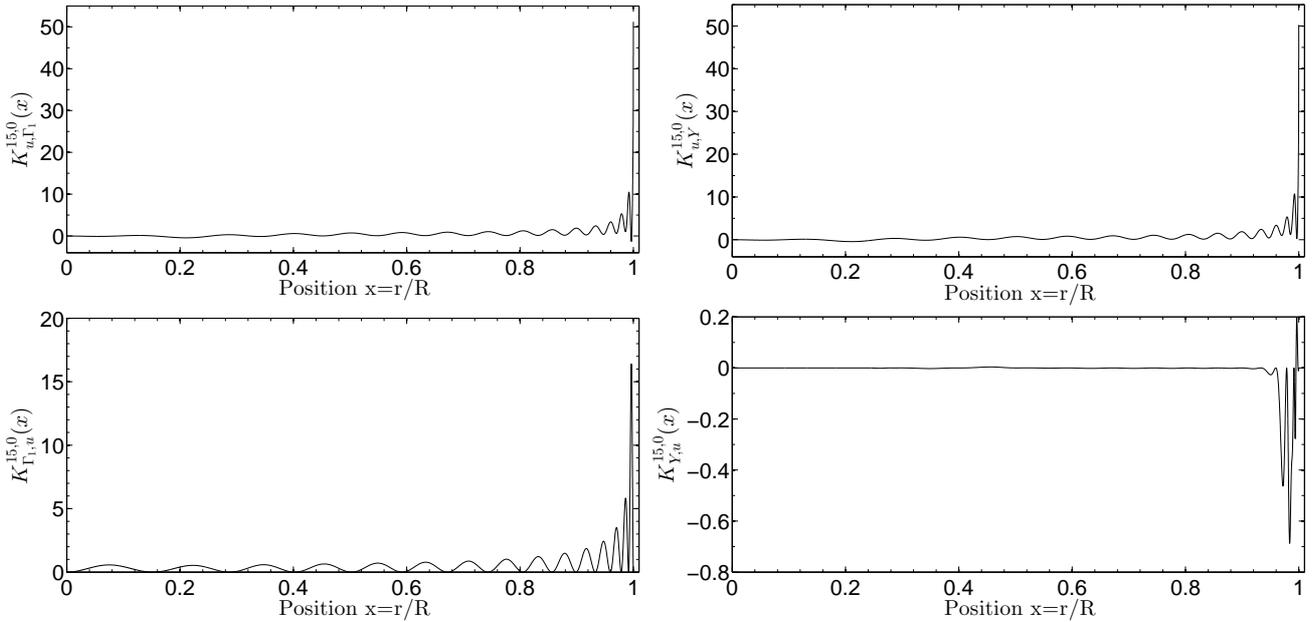}
	\caption{Structural kernels for the $n=15$, $\ell=0$ mode associated with the $(u,\Gamma_{1})$ couple on the left hand side and with the $(u,Y)$ couple on the right hand side.}
		\label{figkeru}
\end{figure*}

As mentioned in the previous section, the method described in \citet{Masters} allows us to derive differential equations for structural kernels. In what follows, we will apply this method to the $(u,\Gamma_{1})$ and the $(u,Y)$ kernels. However, this approach can be applied to many other structural pairs such as: $(c^{2},\Gamma_{1})$, $(c^{2},Y)$, $(g,\Gamma_{1})$, $(g,Y)$, \ldots, with $c^{2}=\frac{\Gamma_{1}P}{\rho}$ the squared adiabatic sound speed , $g=\frac{Gm}{r^{2}}$ the local gravity, $\Gamma_{1}=\left( \frac{\partial \ln T}{\partial \ln P}\right )_{S}$ the adiabatic gradient and $Y$ the local helium abundance. We will not describe these kernels since they are straightforward to obtain using the same technique as what will be used here for the $(u,\Gamma_{1})$ kernels. One should also note that a differential equation cannot be obtained for the couple $(N^{2},c^{2})$, with $N^ {2}$ the Brunt-Väisälä frequency, defined as: $N^{2}=\frac{1}{g}\left(\frac{1}{\Gamma_{1}}\frac{d \ln P}{dr}-\frac{d \ln \rho}{dr} \right)$ , without neglecting a supplementary surface term. 
\\
\\
The first step is to assume that if these kernels exist, they should satisfy an integral equation of the type given in Eq. \ref{eqrhoc2couple}, thereby leading to:
\begin{align}
\frac{\delta \nu_{i}}{\nu_{i}}&=\int_{0}^{1}K^{i}_{\rho,\Gamma_{1}}\frac{\delta \rho}{\rho}dx + \int_{0}^{1}K^{i}_{\Gamma_{1},\rho}\frac{\delta \Gamma_{1}}{\Gamma_{1}}dx, \nonumber \\
&= \int_{0}^{1}K^{i}_{u,\Gamma_{1}}\frac{\delta u}{u}dx + \int_{0}^{1}K^{i}_{\Gamma_{1},u}\frac{\delta \Gamma_{1}}{\Gamma_{1}}dx.
\label{equgammafreq}
\end{align}
From that point, we use the definition of $u$ to express the first integral in terms of a density perturbation. This is done using the definition of the pressure, $P$, and the cumulative mass up to a radial position, $r$:
\begin{align}
P=\int_{r}^{R}\frac{G\tilde{m} \tilde{\rho}}{\tilde{r}^{2}}d\tilde{r}+P_{surf}\label{eqpressure} \\
m=\int_{0}^{r}4 \pi \tilde{r}^{2} \tilde{\rho} d\tilde{r} \label{eqmass}
\end{align}
where we will neglect the pressure perturbation at the surface. In what follows, we will use the non-dimensional forms $\hat{P}=\frac{PR^{4}}{GM^{2}}$, where $M$ is the stellar mass, $R$ the stellar radius and $G$ the gravitational constant, $\hat{m}=\frac{m}{M}$ and $\hat{\rho}=\frac{R^{3}\rho}{M}$. To avoid any confusion in already rather intricated equations, we will drop the hat notation in what follows and denote these non-dimensional variables: $P$, $m$ and $\rho$. Using Eqs. \ref{eqpressure} and \ref{eqmass}, one can relate $u$ perturbations to $P$ and $\rho$ perturbations as follows:
\begin{align}
\frac{\delta u}{u}=\frac{\delta P}{P}-\frac{\delta \rho}{\rho} \label{eqdeltau}
\end{align}
However, using Eq. \ref{eqpressure}, one can also relate $P$ perturbations to $\rho$ perturbations. Doing this, one should note that the surface pressure perturbation is usually neglected and considered as a so-called surface effect. Using non-dimensional variables and combining Eq. \ref{eqpressure} and Eq. \ref{eqmass} in Eq. \ref{eqdeltau}, one obtains an expression relating $u$ perturbations solely to $\rho$ perturbations (of course, this is an integral relation due to the definition of the hydrostatic pressure, $P$). One can use this relation to replace $\frac{\delta u}{u}$ in Eq. \ref{equgammafreq} and after the permutation of the integrals stemming from the definition of the hydrostatic pressure perturbation, one obtains the following integral equation relating $K^{i}_{\rho,\Gamma_{1}}$ to $K^{i}_{u,\Gamma_{1}}$:
\begin{align}
\int_{0}^{1}K^{i}_{\rho,\Gamma_{1}} & \frac{\delta \rho}{\rho}dx + \int_{0}^{1}K^{i}_{\Gamma_{1},\rho}\frac{\delta \Gamma_{1}}{\Gamma_{1}}dx = \int_{0}^{1}\left(\frac{m(x)\rho}{x^{2}}\left[\int_{0}^{x}\frac{K^{i}_{u,\Gamma_{1}}}{\bar{P}}d\bar{x}\right] \right. \nonumber \\ & \left. +4 \pi x^{2} \rho\left[\int_{x}^{1} \frac{\tilde{\rho}}{\tilde{x}^{2}} \left[  \int_{0}^{\tilde{x}}\frac{K^{i}_{u,\Gamma_{1}}}{\bar{P}} d\bar{x} \right] d\tilde{x} \right] -K^{i}_{u,\Gamma_{1}}\right) \frac{\delta \rho}{\rho} dx \nonumber \\ 
&+ \int_{0}^{1}K^{i}_{\Gamma_{1},u}\frac{\delta \Gamma_{1}}{\Gamma_{1}}dx.
\end{align}
One should be careful when solving this equation since one is confronted to multiple integrals, with certain equilibrium variables associated to $\tilde{x}$ or $\bar{x}$. Therefore care should be taken when integrating to check the quality of the result. To obtain a differential equation, we note that it is clear that the equation is satisfied if the integrands are equal, meaning that the kernels are related as follows:
\begin{align}
K^{i}_{\rho,\Gamma_{1}}&= \frac{m(x)\rho}{x^{2}}\left[\int_{0}^{x}\frac{K^{i}_{u,\Gamma_{1}}}{\bar{P}}d\bar{x}\right] \nonumber \\ & +4 \pi x^{2} \rho\left[\int_{x}^{1} \frac{\tilde{\rho}}{\tilde{x}^{2}} \left[  \int_{0}^{\tilde{x}}\frac{K^{i}_{u,\Gamma_{1}}}{\bar{P}} d\bar{x} \right] d\tilde{x} \right] -K^{i}_{u,\Gamma_{1}}, \label{eqkernelugamma} \\ 
K^{i}_{\Gamma_{1},\rho} & = K^{i}_{\Gamma_{1},u}.  \label{eqkernelgammau}
\end{align}
Given this integral expression, one can simply derive and simplify the expression to obtain a second order ordinary differential equation in $x$ as follows:
\begin{align}
-y\frac{d^{2}\kappa^{'}}{(dy)^{2}}&+\left[ \frac{2 \pi y^{3/2}\tilde{\rho}}{\tilde{m}}-3 \right]\frac{d \kappa^{'}}{dy}=y\frac{d^{2} \kappa}{(dy)^{2}} \nonumber\\ 
&-\left[ \frac{2 \pi y^{3/2} \tilde{\rho}}{\tilde{m}}-3 +\frac{\tilde{m} \tilde{\rho}}{2 y^{1/2} \tilde{P}} \right]\frac{d \kappa}{dy} \nonumber\\ &+ \left[ \frac{\tilde{m} \tilde{\rho}}{4 y \tilde{P}^{2}} \frac{d\tilde{P}}{dx} - \frac{\tilde{m}}{4y \tilde{P}} \frac{d \tilde{\rho}}{dx}-\frac{3}{4y^{1/2} \tilde{P}}\frac{d\tilde{P}}{dx} -\frac{\tilde{m} \tilde{\rho}}{2y^{3/2}\tilde{P}}\right] \kappa, \label{eqdiffugamma}
\end{align}
where $\kappa=\frac{K^{i}_{u,\Gamma_{1}}}{x^{2}\rho}$, $\kappa^{'}=\frac{K^{i}_{\rho, \Gamma_{1}}}{x^{2}\rho}$ and $y=x^{2}$. The central boundary condition in terms of $\kappa$ and $\kappa^{'}$ is obtained by taking the limit of Eq. (\ref{eqdiffugamma}) as $y$ goes to $0$. The additional boundary conditions are obtained from Eq. \ref{eqkernelugamma}. Namely, we impose that the solution satisfies Eq. \ref{eqkernelugamma} at some point of the domain. This system is then discretised using a finite difference scheme based on \citet{Reesefinite}, and solved using a direct band-matrix solver.
\\
\\
Two quality checks can be made to validate our solution, the first one being that every kernel satisfies Eq. \ref{eqkernelugamma}, the second one being that they satisfy a frequency-structure relation (As Eq. \ref{equgammafreq}) within the same accuracy as the classical structural kernels $(\rho,\Gamma_{1})$ or $(\rho,c^{2})$. We can carry out the same analysis as in section \ref{secscalingtest}, keeping in mind that the squared isothermal sound speed will also be implicitely rescaled by the inversion since it is proportional to $\frac{M}{R}$, as is the squared adiabatic sound speed, $c^{2}$. The results of this test are plotted in Fig. \ref{figverifkerugam} as well as an example of the verification of the integral equation for the kernel associated with the $\ell=0$, $n=15$ mode.
\\
\\
The equation for the $(u,Y)$ kernels is identical when using the following relation:
\begin{align}
\frac{\delta \nu_{i}}{\nu_{i}}&=\int_{0}^{1}K^{i}_{\rho,Y}\frac{\delta \rho}{\rho}dx + \int_{0}^{1}K^{i}_{Y,\rho} \delta Y dx, \nonumber \\
&= \int_{0}^{1}K^{i}_{u,Y}\frac{\delta u}{u}dx + \int_{0}^{1}K^{i}_{Y,u}\delta Ydx.
\label{equYfreq}
\end{align}
meaning that Eq. \ref{eqdiffugamma} can simply be transposed using the definitions: $\kappa=\frac{K^{i}_{u,Y}}{x^{2}\rho}$ and $\kappa^{'}=\frac{K^{i}_{\rho,Y}}{x^{2}\rho}$. One could also start from Eq. \ref{equgammafreq}, use the following definition:
\begin{align}
\frac{\delta \Gamma_{1}}{\Gamma_{1}} &= \left(\frac{\partial ln(\Gamma_{1})}{\partial ln(P)}\right)_{Z,Y,\rho} \frac{\delta P}{P} + \left(\frac{\partial ln (\Gamma_{1})}{\partial ln(\rho)}\right)_{Z,Y,P} \frac{\delta \rho}{\rho} + \left(\frac{\partial ln (\Gamma_{1})}{\partial Y}\right)_{Z,P,\rho}\delta Y \nonumber \\ & +\left(\frac{\partial ln(\Gamma_{1})}{\partial Z}\right)_{Y,P,\rho} \delta Z, \label{eqgammastate}
\end{align}
and neglect the $\delta Z$ contribution. This assumption is particularly justified if one has spectroscopic constraints on the metallicity. Nevertheless, the term associated with $\delta Z$ is smaller than the three other terms and if one is probing the core regions, the $\frac{\delta \Gamma_{1}}{\Gamma_{1}}$ contribution is already very small. Consequently, all of the terms of Eq. \ref{eqgammastate} are small compared to the integral contribution. Still, this assumption is not completely innocent if one wishes to probe surface regions. When comparing the $\left(\frac{\partial ln(\Gamma_{1})}{\partial Z}\right)_{Y,P,\rho}$ to the $\left(\frac{\partial ln (\Gamma_{1})}{\partial Y}\right)_{Z,P,\rho}\delta Y$, we notice that their amplitude are comparable and that the $\left(\frac{\partial ln(\Gamma_{1})}{\partial Z}\right)_{Y,P,\rho}$ is even often larger. However, we have to consider that it will be multiplied by $\delta Z$, which is much smaller than $\delta Y$. Moreover, the functions are somewhat alike in central regions and thus there will be an implicit partial damping of the $\delta Z$ term when damping the $\delta Y$ contribution if it is in the cross-term of the inversion. One should notice that we can control the importance of this assumption by switching from the $(u,Y)$ kernels to the $(u,\Gamma_{1})$ kernels. Indeed, if the error should be important, the inversion result would be changed by the contribution from the neglected term. In conclusion, in the case of the inversion of $t_{u}$ we will present and use in the next sections, this assumption is justified, but this is not certain for inversions of helium mass fraction in upper layers, for which only numerical tests for the chosen indicator will provide a definitive answer.
\\
\\
Knowing these facts, we can search for the $(u,Y)$ kernels and using the previous developments and starting from the $(\rho,Y)$ kernels, directly obtained from $(\rho,\Gamma_{1})$ or $(\rho,c^{2})$, is more straightforward. One should also note that we assume, by using Eq. \ref{eqgammastate}, that the equation of state is known for the target. As illustrated in Fig. \ref{figverifkeruy}, we again test our solutions by plotting the errors on the integral equation (Eq. \ref{equYfreq}), and by seeing how well our solution for the $\ell=0$, $n=15$ mode verifies Eq. \ref{eqkernelugamma}. The $(u,\Gamma_{1})$ and $(u,Y)$ kernels of this particular mode are illustrated in figure \ref{figkeru}. The kernels associated with $u$ are very similar, except for the surface regions where some differences can be seen. 
\section{Indicator for internal mixing processes and evolutionary stage based on the variations of $u$}\label{secnewindic}
\subsection{Definition of the target function and link to the evolutionary effect} \label{sectargetfunc}
\begin{figure*}[t]
	\centering
		\includegraphics[width=17.5cm]{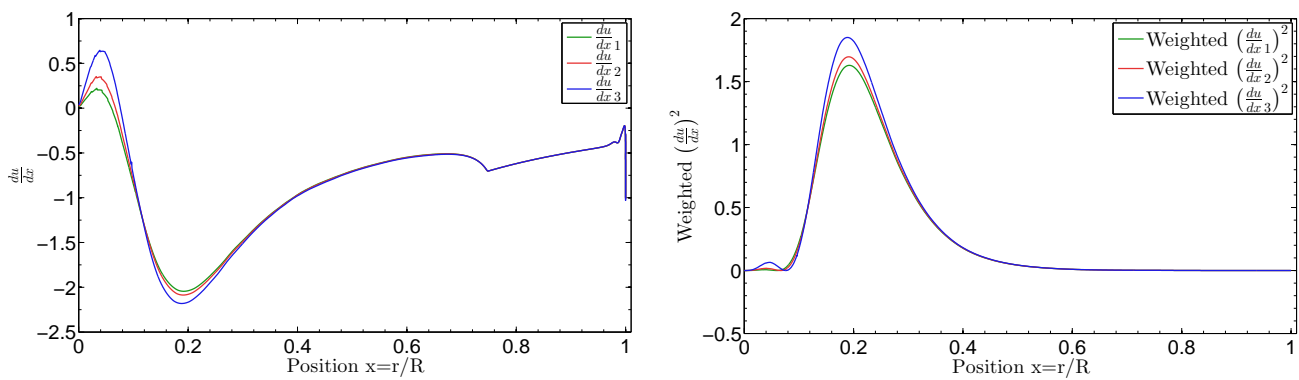}
	\caption{Left panel: structural changes in the scaled $\frac{du}{dx}$ profile with the evolution. The models are main sequence $1.0M_{\odot}$ models. Right panel: the same effects as seen when $\left( \frac{du}{dx} \right)^{2}$ is weighted according to Eq. \ref{eqdefwheightfunc}.}
		\label{figstrucefsens}
\end{figure*}

Knowing that it is possible to obtain the helium abundance in the integral equation \ref{equYfreq}, one could be tempted to use it to obtain corrections on the helium abundance in the core and therefore having insights on the chemical evolution directly. However, Fig. \ref{figkeru} reminds us of the hard reality associated with these helium kernels. Their intensity is only non-negligible in surface regions and it is thus impossible to obtain information on the core helium abundance using such kernels. 
\\
\\
Another approach would be to use the squared isothermal sound speed, $u$, to reach our goal. Indeed, we know that $u=\frac{P}{\rho}\propto \frac{T}{\mu}$, and that during the evolution along the main sequence, the mean molecular weight will change. Moreover the core contraction can also lead to changes in the variation of the $u$ profile. Using the same philosophy as for the definition of the first age indicator \citep[see][]{Buldgen} and ultimately for the use of the small separation as an indicator of the core conditions \citep[see][]{Tassoul}, we build our indicator using the first radial derivative of $u$. Using $u$ instead of $c^{2}$ allows us to avoid the dependence in $\Gamma_{1}$ which is responsible for the surface dependence of $\frac{dc}{dr}$. To build our indicator, we analyse the effect of the evolution on the profile of $\frac{du}{dr}$. This effect is illustrated in Fig. \ref{figstrucefsens}. As can be seen, two lobes tend to develop as the star ages. The first problem is that these variations are of opposite signs, meaning that if we integrate through both lobes, the sensitivity will be greatly reduced. Therefore, we choose to base our indicator on the squared first radial derivative:
\begin{align}
\bar{t}_{u}=\int_{0}^{R}f(r)\left(\frac{du}{dr}\right)^{2}dr \label{eqdefindic},
\end{align}
where $f(r)$ is an appropriate weight function. First, we will consider that the observed star and the reference model have the same radius. The target function for this indicator can easily be obtained. We perturb the equation for $t_{u}$ and use an integration by parts to relate the perturbations of the indicator to structural perturbations of $u$:
\begin{align}
\frac{\delta \bar{t}_{u}}{\bar{t}_{u}}&=\frac{2}{\bar{t}_{u}}\int_{0}^{R}f(r)\frac{du}{dr}\frac{d \delta u}{dr}dr, \nonumber \\
&=\frac{-2}{\bar{t}_{u}}\int_{0}^{R} u \frac{d}{dr}\left( f(r)\frac{du}{dr} \right) \frac{\delta u}{u} dr + \left[ f(r)\frac{du}{dr} \delta u \right]_{0}^{R}.   
\end{align}

The last term on the second line is not suitable for SOLA inversions, given the neglect of surface terms in the kernels. Hence we define the function $f$ so that $f(0)=f(R)=0$, thereby cancelling this term. This leads to the following expression:
\begin{align}
\frac{\delta \bar{t}_{u}}{\bar{t}_{u}}=\int_{0}^{R}\mathcal{T}_{t_{u}}\frac{\delta u}{u}dr,
\label{eqperturbt}
\end{align}
where

\begin{align}
\mathcal{T}_{t_{u}}=\frac{-2 u}{\bar{t}_{u}} \frac{d}{dr}\left( f(r)\frac{du}{dr} \right). \label{eqtargetfuncage}
\end{align}
The weight function $f(r)$ must be chosen according to a number of criteria: it has to be sensitive to the core regions where the profile changes; it has to be of low amplitude at the boundaries of the domain, allowing us to do the integration by parts necessary to obtain $\delta u$ in the expression; it should be possible to fit the target function associated with this $f(r)$ using structural kernels from a restricted number of frequencies; moreover, Eq. (\ref{eqtargetfuncage}) being related to linear perturbations, it is clear that non-linear effects should not dominate the changes of this indicator\footnote{Otherwise, using the SOLA technique, which is linear, would be impossible.}. We also know that the amplitude of the structural kernels is $0$ in the centre, so $f(r)$ should also satisfy this condition.
\\
\\
We define the weight function as follows:
\begin{align}
f(r)= r^{\alpha} \left( r-R \right)^{\lambda} \exp \left(-\gamma \left(\frac{r-r_{opt}}{R}\right)^{2}\right), \label{eqdefwheightfunc}
\end{align}
which means that we have $4$ parameters to adjust. The case of $r_{opt}$ is quickly treated. Since we know that the changes will be localised in the lobes developing in the core regions, we chose to put $r_{opt}=0$. $\alpha$ and $\lambda$ should be at least $1$, so that the integration by part is exact and the central limit for the target function and the structural kernels is the same. $\gamma$ depends on the effects of the non-linearities. However, since we will have to perform a second derivative of $u$ a more practical concern appears: we don't want to be influenced by the effects of the discontinuity at the boundary of the convective envelope. At the end of the day, we use the following set of parameters: $\alpha=1$, $\lambda=2$, $\gamma=7$  and $r_{opt}=0$. One could argue that the optimal choice for $r_{opt}$ would be at the maximum of the second lobe or between both lobes to obtain the maximal sensitivity in the structural variations. These values were also tested, but the results were a little less accurate than using $r_{opt}=0$ and they involved higher inversion coefficients and therefore higher error magnification. We illustrate the weighted profile obtained for this optimal set of parameters in the right-hand side panel of Fig. \ref{figstrucefsens}. Furthermore, Eq. \ref{eqperturbt} is satisfied up to $5 \%$, so we can try to carry out inversions for this indicator. It is also important to note that for the sake of simplicity, we do not chose to change the values of these parameters with the model, which would only bring additional complexity to the problem.
\\
\\
Thanks to the target function $\mathcal{T}_{t_{u}}$, we can now carry out inversions for the integrated quantity $t_{u}$ using the linear SOLA inversion technique \citep{Pijpers}. First, let us take some time to recall the purpose of inversions and our adaptation of the SOLA technique to integrated quantities. Historically, inversions have been used to obtain seismically constrained structural profiles \citep{Basudiff} as well as rotation profiles \citep{Schou} in helioseismology. However, all these methods are not well suited for the inversions we wish to carry here. As discussed in \citet{Buldgen}, the SOLA inversion technique, which uses a ``kernel matching'' approach is well suited to our purpose. Indeed, this approach allows us to define custom-made target functions that will be used to build a cost function, here denoted $\mathcal{J}$. In the case of the $t_{u}$ quantity, one has the following definition:
\begin{align}
\mathcal{J}_{t_{u}} = &\int_{0}^{1}\left[ K_{\mathrm{Avg}}-\mathcal{T}_{t_{u}}\right]^{2}dx +\beta \int_{0}^{1}K^ {2}_{\mathrm{Cross}}dx + \tan(\theta) \sum^{N}_{i}(c_{i}\sigma_{i})^{2} \nonumber \\
 &+ \eta \left[ \sum^{N}_{i}c_{i}-k \right]
\end{align}
where  $K_{\mathrm{Avg}}$ is the so-called averaging kernel and $K_{\mathrm{Cross}}$ is the so-called cross-term kernel defined as follows for the $(u,\Gamma_{1})$ structural pair (For $(u,Y)$, one would replace $K^{i}_{u,\Gamma_{1}}$ with $K^{i}_{u,Y}$ and $K^{i}_{\Gamma_{1},u}$ with $K^{i}_{Y,u}$):
\begin{align}
K_{\mathrm{Avg}}=\sum_{i}^{N}c_{i}K^{i}_{u,\Gamma_{1}}, \\
K_{\mathrm{Cross}}=\sum_{i}^{N}c_{i}K^{i}_{\Gamma_{1},u}.
\end{align}
$\theta$ and $\beta$ are free parameters of the inversion. $\theta$ is related to the compromise between the amplification of the observational error bars $(\sigma_{i})$ and the fit of the kernels, whereas $\beta$ is allowed to vary to give more weight to the elimination of the cross-term kernel. In this expression, $N$ is the number of observed frequencies, the $c_{i}$ are the inversion coefficients, which will be used to determine the correction to be applied on the $t_{u}$ value. $\eta$ is a Lagrange multiplier and the last term appearing in the expression of the cost-function is a supplementary constraint applied to the inversion which is presented in Sect. \ref{Nonlinear}.
\\
\\
If the observed target and the reference model have the same radius, the inversion will measure the value of $t_{u}$ for the observed target. However, if this condition is not met, the inversion will produce a \textit{scaled} value of this indicator. By defining integral equations such as Eq. \ref{eqdefindic}, or even Eq. \ref{eqrhoc2couple}, we have seen in Sect. \ref{secscalingtest} that we made the hypothesis that both target and reference model had the same radius. However, as the frequencies scale with $\left(\frac{M}{R^{3}}\right)^{1/2}$, the inversion will preserve the mean density of the observed target. Therefore, we are implicitly carrying out the inversion for a \textit{scaled} target homologous to the observed target, which has the radius of the reference model but the mean density of the observed target. Simple reasoning shows us that the mass of this scaled target is: $\bar{M}_{tar}=M_{tar}\frac{R^{3}_{ref}}{R_{tar}^{3}}$. Thus, as $t_{u}$ scales as $M^{2}$, there is a difference between the target value $t_{u}^{obs}$ and the measured value, $t_{u}^{inv}$. We can write the following equations:
\begin{align}
\frac{t_{u}^{inv}}{\bar{M}_{tar}^{2}}&=\frac{t_{u}^{obs}}{M^{2}_{tar}}, \\
\frac{t_{u}^{inv}}{R_{ref}^{6}}&=\frac{t_{u}^{obs}}{R_{tar}^{6}} \label{eqcriterion},
\end{align}
where we have used the definition of $\bar{M}_{tar}$ to express the mass dependencies as radius dependencies. Therefore, we will use Eq. \ref{eqcriterion} as a criterion to determine whether the inversion was successful or not. 
\subsection{Non-linear generalisation}\label{Nonlinear}
We present in this section a general approach to the non-linear generalisation presented in \citet{Reese} and \citet{Buldgen} for any type of global characteristic that can scale with the mass of the star. We can say that the frequencies scale as $M^{1/2}$. Of course, they scale as the mean density, namely $\sqrt{\frac{M}{R^{3}}}$. However, since the inversion works with a fixed radius and implicitely scales the target to the same radius as the reference model, the iterative process associated with this non-linear generalisation will never change the model radius. Therefore, we do not take this dependence into account and simply work on the mass dependence. Let $A$ be a global characteristic, related to the mass of the star. It is always possible to define a factor $k$ so that:
\begin{align}
A\propto \nu^{k}.
\end{align}
And we obtain directly:
\begin{align}
\frac{\delta A}{A}=k\frac{\delta \nu}{\nu}.
\end{align}
However, using the definition of the inverted correction of $A$, one has:
\begin{align}
\frac{\delta A}{A} = \sum_{i}c_{i}\frac{\delta \nu_{i}}{\nu_{i}},
\end{align}
with the $c_{i}$ the inversion coefficients. Using the same reasoning as in \citet{Buldgen} and \citet{Reese}, we define the inversion as ``unbiased'' (term which should not be taken in the statistical sense.) if it satisfies the condition:
\begin{align}
\sum_{i}c_{i}=k
\end{align}
Now we can define an iterative process, using the scale factor $s^{k}_{0}=\frac{A_{\mathrm{inv},0}}{A_{\mathrm{ref}}}$, we will scale the reference model (in other words, multiply its mass and density by $s^ {2}$, its pressure by $s^{4}$, leaving $\Gamma_{1}$ unchanged) and carry out a second inversion, then define a new scale factor and so on until no further correction is made by the inversion process. In other words, we search for the fixed point of the equation of the scale factor. At the $j^{th}$ iteration, we obtain the following equation for the inversion value of $A$:
\begin{align}
A_{\mathrm{inv}}&=A_{\mathrm{ref}}s^{k-1}_{j}\left[ s_{j}+\sum_{i}c_{i}\frac{\delta \nu_{i}}{\nu_{i}} +k(1-s_{j}) \right]. 
\end{align}
This can be written in terms of the scale factor only, noting that at the $j^{th}$ iteration, $\frac{A_{inv}}{A_{ref}}=s^{k}_{j+1}$:
\begin{align}
s^{k}_{j+1}=s_{j}^{k-1}\left[ s_{j}+\sum_{i}c_{i}\frac{\delta \nu_{i}}{\nu_{i}} +k(1-s_{j}) \right]. 
\end{align}
The fixed point is then easily obtained:
\begin{align}
s_{\mathrm{opt}}=\frac{\sum_{i}c_{i}\frac{\delta \nu_{i}}{\nu_{i}}}{k}+1,
\end{align}
and can be used directly to obtain the optimal value of the indicator $A$. We can also carry out a general analysis of the error bars treating the observed frequencies and $A_{\mathrm{inv}}$ as stochastic variables:
\begin{align}
\nu_{\mathrm{obs},i}=\bar{\nu}_{\mathrm{obs},i}(1+\epsilon_{i}), \\
A_{\mathrm{inv}}=\bar{A}_{\mathrm{inv}}(1+\epsilon_{A}),
\end{align} 
with $\epsilon_{i}$ and $\epsilon_{A}$ being the stochastic contributions to the variables. Using the hypothesis that $\epsilon_{i}\ll 1$ we easily obtain the following equation for the error bars:
\begin{align}
\sigma_{A}=A_{ref}\left[ \frac{1}{k} \sum_{i}c_{i}\frac{\nu_{i}^{\mathrm{obs}}}{\nu^{\mathrm{ref}}_{i}} \right]^{k-1}\sqrt{(\sum_{i}c_{i}\sigma_{i})^{2}}.
\end{align}
We note that in the particular case of the indicator $t_{u}$, $k=4$. Indeed, $t_{u}\propto M^{2}$ whereas $\nu \propto M^{1/2}$.
\subsection{Tests using various physical effects}\label{secphystest}
\begin{figure*}[t]
	\centering
		\includegraphics[width=17.5cm]{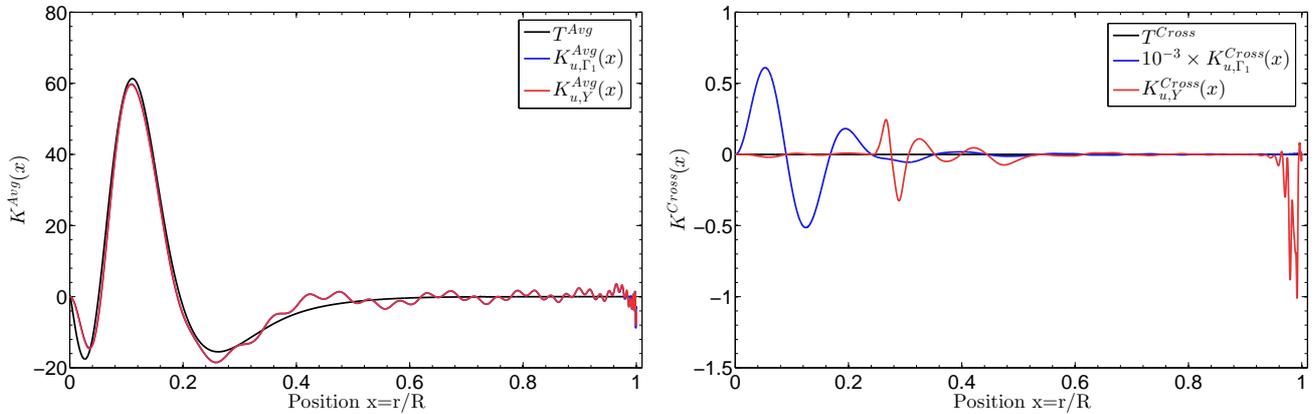}
	\caption{The left hand-side figure illustrates the averaging kernels for both structural pairs and their respective targets. The red and blue curves are nearly identical, so that only the red curve is visible. The right hand-side figure illustrates the cross-term kernels for both structural pairs, the target being $0$.}
		\label{figkerage}
\end{figure*}

To test the accuracy of the SOLA technique applied to the $t_{u}$ indicator, we carried out the same test as in \citet{Buldgen} using stellar models which would play the role of observed targets. These models included physical phenomena not taken into account in the reference models. A total of $13$ targets were constructed, with masses of $0.9$ $M_{\odot}$, $1.0$ $M_{\odot}$ and $1.1$ $M_{\odot}$ but to avoid redundancy, we only present $6$ of them which are representative of the mass and age ranges and of the physical effects considered in our study. We tested various effects for each mass, namely those that come from microscopic diffusion using the approach presented in \citet{Thoul} (multiplying the atomic diffusion coefficients by a factor given in the last line of Table \ref{tabcaractarg}\footnote{These values might seem excessive regarding the reliability of the implementation of diffusion. We stress here that our goal was to witness the impact of significant changes on the results. However, other processes or mismatches could alter the $\frac{du}{dr}$ gradient and thus be detected by the inversion.}), those caused by a helium abundance mismatch, those that result from a metallicity mismatch, as well as those which stem from using a different solar heavy elements mixture. For the last case, the target was built using the Grevesse-Noels $1993$ (GN$93$, \citet{GrevesseNoels}) abundances  and the reference model was built using the Asplund-Grevesse-Sauval-Scott $2009$ (AGSS $2009$, \citet{AGSS}) abundances. All targets and reference models were built using version $18.15$ of the Code Liégeois d'Evolution Stellaire (CLES) stellar evolution code \citep{ScuflaireCles} and their oscillations frequencies were calculated using the Liège oscillation code (LOSC, \citet{ScuflaireLosc}). Table \ref{tabcaractarg} summarises the properties of the $6$ targets presented in this paper.
\begin{table*}[t]
\caption{Characteristics of the $6$ targets. The differences between target and reference model are given in boldface.}
\label{tabcaractarg}
  \centering
\begin{tabular}{r | c | c | c | c | c | c}
\hline \hline
 & \textbf{Target$_{1}$}& \textbf{Target$_{2}$}& \textbf{Target $_{3}$}&\textbf{Target$_{4}$}& \textbf{Target$_{5}$}& \textbf{Target$_{6}$}\\ \hline
 \textit{Mass ($\mathrm{M_{\odot}}$)}& $1.0$ & $1.0$ & $1.0$ & $0.9$ &  $1.0$ &$1.1$ \\
\textit{Radius ($\mathrm{R_{\odot}}$)} & $1.076$ & $1.159$&$1.14$&$0.89$&$1.193$&$1.297$\\ 
\textit{Age ($\mathrm{Gyr}$)} &$8.05$ &$7.55$&$7.06$&$6.0$&$5.121$& $5.135$ \\ 
\textit{$T_{\mathrm{eff}}$ ($\mathrm{K}$)} & $5597$ & $5884$&$5712$&$5329$&$6081$&$5967$ \\ 
\textit{$Z_{0}$} & $\mathbf{0.015}$ & $0.0122$&$0.0122$&$0.0122$&$0.0122$&$0.0122$ \\
\textit{$Y_{0}$} & $0.2457$ & $0.2485$&$0.2485$&$0.2485$&$\mathbf{0.3078}$&$0.2485$ \\
\textit{Abundances} & AGSS09 & \textbf{GN93} & AGSS09 & AGSS09 & AGSS09 & AGSS09 \\
\textit{$\alpha_{\mathrm{MLT}}$} & $1.522$ & $1.522$ & $1.522$ & $1.522$ & $1.522$&$1.522$ \\
\textit{Diffusion factor} & $0$ & $0$ &$\mathbf{2}$&$\mathbf{1.6}$&$0$&$\mathbf{1.6}$\\
\hline
\end{tabular}
\end{table*}
The selection of the reference model was based on the fit of the large and small separation for $60$ modes with $n=7-26$ and $\ell=0-2$ using a Levenberg-Marquardt minimization code. The use of supplementary constraints will be discussed in Sect. \ref{secmodelstudy} whereas the effects of the selection of the modes will be discussed in Sect. \ref{secmodestudy}. The choice of $60$ frequencies is motivated by the number of observed frequencies for the system $16$Cyg $A$ - $16$Cyg $B$ by Kepler, which is between $50$ and $60$ \citep{Verma}. The inversions were carried out using the $(u,Y)$ and the $(u,\Gamma_{1})$ structural kernels. 
\\
\\
If the inversion of $t_{u}$ shows that there are differences between the target and the reference model, then we know that the core regions are not properly represented. Whether these differences arise from atomic diffusion or helium abundance mismatch, the $t_{u}$ indicator alone could not answer this question\footnote{Constraining properly the changes due to multiple additional mixing processes with only one structural indicator is of course impossible.}. Therefore, the philosophy we adopt in this paper is simply the following: "Is the inversion able to correct mistakes in the reference models? If yes, within what accuracy range?". The capacity of disentangling between different effects is partially illustrated in Sect. \ref{secmodelstudy}, but additional indicators are still required to provide the best diagnostic possible given a set of frequencies. 
\\
\\
The results are given in Table \ref{tabresultsuGam} for the $(u,\Gamma_{1})$ kernels and Table \ref{tabresultsuY} for the $(u,Y)$ kernels along with the respective error contributions given according to the developments of \citet{Reese} and \citet{Buldgen}. We denote these error contributions: $\varepsilon_{\mathrm{Avg}}$, $\varepsilon_{\mathrm{Cross}}$, $\varepsilon_{\mathrm{Res}}$. These errors contributions being defined as follows:
\begin{align}
\varepsilon_{\mathrm{Avg}}=\int_{0}^{1}\left[ K_{\mathrm{Avg}}-\mathcal{T}_{t_{u}} \right]\frac{\delta u}{u}dx, \\
\varepsilon_{\mathrm{Cross}}=\int_{0}^{1} K_{\mathrm{Cross}}\frac{\delta \Gamma_{1}}{\Gamma_{1}}dx,
\end{align}
if the $(u,\Gamma_{1})$ couple is used. If one prefers the $(u,Y)$, $\varepsilon_{\mathrm{Cross}}$ becomes:
\begin{align}
\varepsilon_{\mathrm{Cross}}=\int_{0}^{1}K_{\mathrm{Cross}}\delta Y dx.
\end{align}
Finally, $\varepsilon_{\mathrm{Res}}$ is associated to the residual contribution, in the sense that it is what remains after one has taken into account both $\varepsilon_{\mathrm{Cross}}$ and $\varepsilon_{\mathrm{Avg}}$. The target function and their fits are illustrated in Fig. \ref{figkerage} for $\mathrm{Target}_{4}$. 
\begin{table*}[t]
\caption{Inversion results for the $6$ targets using the $(u, \Gamma_{1})$ kernels.}
\label{tabresultsuGam}
  \centering
\begin{tabular}{r | c | c | c | c | c | c }
\hline \hline
 & $\frac{t_{u}^{\mathrm{ref}}}{R_{ref}^{6}}$ $(g^{2}/cm^{6})$ & $\frac{t_{u}^{\mathrm{inv}}}{R_{ref}^ {6}}$ $(g^{2}/cm^{6})$ $(u,\Gamma_{1})$ & $\frac{t_{u}^ {\mathrm{obs}}}{R_{tar}^ {6}}$ $(g^{2}/cm^{6})$& $\varepsilon_{\mathrm{Avg}}^{u,\Gamma_{1}}$& $\varepsilon_{\mathrm{Cross}}^{u,\Gamma_{1}}$&$\varepsilon_{\mathrm{Res}}^{u,\Gamma_{1}}$\\ \hline
 $\mathrm{Target}_{1}$& $4.032$ & $3.568\pm 0.063$ & $3.532$ & $4.415\times 10^{-4}$& $-1.684\times 10^{-4}$& $9.767\times 10^{-3}$\\
 $\mathrm{Target}_{2}$ & $3.434$ &$3.24\pm 0.075$&$3.428$& $-1.301 \times 10^{-3}$& $-4.419 \times 10^{-4}$ & $5.951 \times 10^{-4}$\\ 
 $\mathrm{Target}_{3}$ &$3.562$ &$3.275\pm 0.067$&$3.252$& $5.789 \times 10^{-3}$& $-1.178 \times 10^{-3}$& $2.277 \times 10^{-3}$ \\ 
  $\mathrm{Target}_{4}$ & $5.879$&$5.621\pm 0.147$&$5.536$&$1.388 \times 10^{-2}$& $3.088 \times 10^{-4}$&$8.062 \times 10^{-4}$ \\
 $\mathrm{Target}_{5}$ & $2.845$ &$2.669\pm 0.088$&$2.630$& $-6.498 \times 10^{-4}$& $-4.493\times 10^{-3}$& $2.366\times 10^{-4}$\\
  $\mathrm{Target}_{6}$ & $3.205$ &$3.480\pm 0.091$&$3.498$& $1.496 \times 10^{-2}$& $-1.11\times 10^{-3}$& $7.824\times 10^{-4}$\\
\hline
\end{tabular}
\end{table*}
\begin{table*}[t]
\caption{Inversion results for the $6$ targets using the $(u,Y)$ kernels.}
\label{tabresultsuY}
  \centering
\begin{tabular}{r | c | c | c | c | c | c }
\hline \hline
 & $\frac{t_{u}^{\mathrm{ref}}}{R_{ref}^{6}}$ $(g^{2}/cm^{6})$ & $\frac{t_{u}^{\mathrm{inv}}}{R_{ref}^ {6}}$ $(g^{2}/cm^{6})$ $(u,Y)$ & $\frac{t_{u}^ {\mathrm{obs}}}{R_{tar}^ {6}}$ $(g^{2}/cm^{6})$& $\varepsilon_{\mathrm{Avg}}^{u,Y}$& $\varepsilon_{\mathrm{Cross}}^{u,Y}$&$\varepsilon_{\mathrm{Res}}^{u,Y}$\\ \hline
 $\mathrm{Target}_{1}$& $4.032$ & $3.575\pm 0.063$ & $3.532$ & $8.34\times 10^{-4}$& $1.601\times 10^{-4}$& $1.1\times 10^{-2}$\\
 $\mathrm{Target}_{2}$ & $3.434$ &$3.423\pm 0.075$&$3.428$& $-1.301 \times 10^{-3}$& $1.338 \times 10^{-7}$ & $2.127 \times 10^{-5}$\\ 
 $\mathrm{Target}_{3}$ &$3.562$ &$3.283\pm 0.067$&$3.252$& $5.748 \times 10^{-3}$& $8.296 \times 10^{-3}$& $-4.794 \times 10^{-3}$ \\ 
  $\mathrm{Target}_{4}$ & $5.879$&$5.624\pm 0.148$&$5.536$&$1.386 \times 10^{-2}$& $1.337 \times 10^{-3}$&$5.448 \times 10^{-4}$ \\
 $\mathrm{Target}_{5}$ & $2.845$ &$2.675\pm 0.089$&$2.630$& $-5.421 \times 10^{-4}$& $-8.184\times 10^{-3}$& $3.168\times 10^{-4}$\\
  $\mathrm{Target}_{6}$ & $3.205$ &$3.480\pm 0.091$&$3.498$& $1.458 \times 10^{-2}$& $1.214\times 10^{-2}$& $-9.721\times 10^{-3}$\\
\hline
\end{tabular}
\end{table*}
As we can see from Tables \ref{tabresultsuGam} and \ref{tabresultsuY}, we obtain rather accurate results for all cases. This means that the inversion is succesfull and that the regularization process is sufficient for the values $\beta=10^{-6}$ and $\theta=10^{-5}$. 
\\
\\
We see in the fifth column of Tables \ref{tabresultsuGam} and \ref{tabresultsuY} that the averaging kernel fit is usually the dominant error contribution. We will see in the next sections how this result changes with the modes used or with the quality of the reference model. If we analyse the cross-term error contribution, we see that it is generally way less important than the averaging kernel mismatch error. We also see that despite the high amplitude of the cross-term kernels associated with $\Gamma_{1}$ shown in Fig. \ref{figkerage}, the real error is rather small and often smaller than the error associated with the helium cross-term kernel. This is due not only to the small variations in $\Gamma_{1}$ between target and reference model but also to the oscillatory behaviour of the $\Gamma_{1}$ cross-term kernel. In contrast, the cross-term kernel of the $(u,Y)$ kernel has a smaller amplitude, but nearly no oscillatory behaviour and is more important in the surface regions, where the inversion is naturally less robust. Nonetheless, the results of the $(u,Y)$ kernels also show some compensation. We also note that they tend to have more important residual errors. However, there is no clear difference in accuracy between the $(u,\Gamma_{1})$ and the $(u,Y)$ kernels. In the case of $\mathrm{Target}_{2}$, we see that although the results are slightly improved, the reference value is within the errors bars of the inversion results. In an observed case, this would mean that the reference model is already very close to the target as far as the indicator $t_{u}$ is concerned. However, we wish to point out that it seems rather improbable that the only difference between a static model and a real observed star would be in its heavy elements mixture.
\\
\\
Analysing the residual contribution is slightly more difficult, since it includes any supplementary effect: surface terms, non-linear contributions, errors in the equation of state (when using kernels related to $Y$), ... We can see in this study that the residual error is rather constricted. This will not be the case for example if the parameter $\theta$ is chosen to be very small, or if the scaling effect has not been taken into account\footnote{In which case one would be searching for a result impossible to obtain.}. In fact, the $\theta$ parameter is a regularizing parameter in the sense that it does not allow the inversion coefficients to take on extremely large values. If that would be the case, the inversion would be completely unstable because a slight error on the fit would be amplified and would lead to wrong results. This is quickly understood knowing that the inversion coefficients are used to recombine the frequencies as follows:
\begin{align}
\frac{\delta t_{u}^{inv}}{t_{u}}=\sum_{i}^{N}c_{i}\frac{\delta \nu_{i}}{\nu_{i}},
\end{align}
with $N$ the total number of observed frequencies. Where this equation is subject to the uncertainties in Eq. \ref{eqperturbt}  and Eq. \ref{equgammafreq} (or, respectively Eq. \ref{equYfreq}), any error will be dramatically amplified by the inversion process. Therefore there is no gain in reducing $\theta$ since at some point, the uncertainties behind the basic equation of the inversion process will dominate and lead the method to failure. In this case, the inversion problem is not sufficiently regularized. Such an example is presented and analysed in the following section.

\section{Impact of the type and number of modes on the inversion results}\label{secmodestudy}

When carrying out inversions on observed data, we are limited to the observed modes. Therefore the question on how the inversion results depend on the type of modes is of uttermost importance. The reason behind this dependence is that different frequencies are associated with different eigenfunctions, in other words different structural kernels, sensitive to different regions of the star. Therefore, the inverse problem will vary for each set of modes because the physical information contained in the observational data changes. Hence, we studied $4$ targets using $7$ sets of modes. As in Sect. \ref{secphystest}, we wish to avoid redundancy and will present our results for one target, namely $\mathrm{Target}_{3}$, and $5$ different frequency sets. As a supplementary test case, we defined two target models with the properties of $16$CygA and $16$CygB found in the litterature \citep{Metcalfe,White, Verma}. Using these properties, we added strong atomic diffusion to the $16$CygA model and used $Z=0.023$ as well as the GN$93$ mixture for $16$CygB. In constrast, the reference models used $Z=0.0122$ and the AGSS$09$ mixture. The characteristics of these models are also summarized in Table \ref{tabcaracsupltarg}. We used the set of observed modes given by \citet{Verma} and ignored the isolated $\ell=3$ modes for which there was no possibility to define a large separation.
\\
\\
All the sets used for the test cases of this section are summarized in Table \ref{tabmodes}. The reference model was chosen as in Sect. \ref{secphystest}, using the arithmetic average of the large and small frequency separations as constraints for its mass and age. Using these test cases, we first analyse whether the inversion results depend on the values of the radial order $n$ of the modes, with the help of the frequency sets $1$, $2$ and $3$ (see Table \ref{tabmodes}). Then we analyse the importance of the $\ell=3$ modes for the inversion using set $4$ and $5$.
\\
\\
The inversion results for all these targets and sets of mode are presented in Tables \ref{tabresultsuGam3modes} and \ref{tabresultsuGamCyg} for the $(u,\Gamma_{1})$ kernels. A first conclusion can be drawn from the results using sets $1$, $2$ and $3$: low $n$ are important to ensure an accurate results. Indeed, set $2$ provides much better results than set $1$, and even using set $3$ (which is only set $1$ extended up to $n=34$ for each $\ell$) does not improve the results any further. This means that modes with $n>27$ are barely used to fit the target.
\\
\\
This first result can be understood in different ways. First, using mathematical reasoning, we can say that the kernels associated with higher $n$ have high amplitudes in the surface regions and are therefore not well suited to probing central regions. Another way of understanding this problem is that when we use modes with high $n$, we come closer to the asymptotic regime and the eigenfunctions are described by the JWKB approximation, all having a similar form and therefore not providing useful additional information. In that sense, there is a clear difference between inverted structural quantities and the information deduced from asymptotic relations, which requires high $n$ values to be valid, thereby highlighting the usefulness of inversions.
\begin{table*}[t]
\caption{Characteristics of the $16$CygA and $16$CygB clones.}
\label{tabcaracsupltarg}
  \centering
\begin{tabular}{r | c | c | c | c | c | c | c}
\hline \hline
 & \textbf{$16$CygA-clone}& \textbf{$16$CygB-clone}\\ \hline
 \textit{Mass ($\mathrm{M_{\odot}}$)}& $1.11$ & $1.06$ \\
\textit{Radius ($\mathrm{R_{\odot}}$)}& $1.13$ & $1.04$ \\ 
\textit{Age ($\mathrm{Gyr}$)} &$6.9$&$6.7$\\ 
\textit{$T_{\mathrm{eff}}$ ($\mathrm{K}$)} & $5696$ & $5772.9$ \\ 
\textit{$Z_{0}$} & $0.024$&$0.023$\\
\textit{$Y_{0}$} &$0.241$&$0.242$\\
\textit{Abundances} & AGSS09 & GN93\\
\textit{$\alpha_{\mathrm{MLT}}$} & $2.0$ & $2.0$ \\
\textit{Diffusion coefficient} & $2$ &$0$\\
\hline
\end{tabular}
\end{table*}
The question of the importance of the modes $\ell=3$ is also quickly answered from the results obtained with set $4$ and $5$. For those test cases, we reach a very good accuracy even without $n\leq 9$, unlike the previous test case using set $2$. Moreover, we use even less frequencies than for the $3$ first sets. This is in fact crucial to determine whether one can apply an inversion in an observed case, since a few $\ell=3$ modes can change the results and make the inversion successful.
\\
\\
To further illustrate the importance of the octupole modes, we use the $16$CygA and $16$CygB clones and carry out inversions for their respective observed frequency sets. First, we use all frequencies and reach a reasonable accuracy for both targets. In the second test case, we do not use the octupole modes and one can observe a drastic change in accuracy. These results are illustrated in Table \ref{tabresultsuGamCyg}, where the notation "Small" (for small frequency set) has been added to the lines associated with the results obtained without using the $\ell=3$ modes.
\\
\\
Looking again at the results for $16$CygA, we again see that although the inversion improves the value of $t_{u}$, the reference value lies within the observational error bars of the inverted result. The case of the truncated set of frequencies is even worse, since the inverted result is less accurate than the reference value. We therefore analysed the problem for the full frequency set. To do so we carried out different inversions using higher values of $\theta$, the results for $\theta=10^{-4}$ are illustrated in Table \ref{tabresultsuGamCyg}. In this case we have smaller error bars, but what is reassuring is that the result did not change drastically when we changed $\theta$. This means that the problem is properly regularized around $\theta=10^{-5}$ and $\theta=10^{-4}$ and that we can trust the inversion results. Our advice is therefore to always look at the behaviour of the solution with the inversion parameters to see if there is any sign of compensation or other undesirable behaviour. In fact there is no law to select the value of $\theta$ and applying fixed values blindly for all asteroseismic observations is probably the best way to obtain unreliable results.
\\
\\
The case of the small frequency set is even more intriguing since the result improves greatly with $\theta=10^{-4}$. The question that arises is whether the problem is not properly regularized with $\theta=10^{-5}$ or whether we are facing some fortuitous compensation effect that leads to very accurate results. If we are facing fortuitous compensation, taking $\theta$ slightly larger than $10^{-4}$ or increasing $\beta$ will drastically reduce the accuracy since any change in the linear combination will affect the compensation. However, if we are facing a regularization problem, the accuracy should decrease regularly with the change of parameters (since we are slightly reducing the quality of the fit with those changes). We emphasize again that one should not choose values of the inversion parameters for which any small augmentation of the regularization would drastically change the result. In this particular case, we were confronted with insufficient regularisation and choosing $\theta=10^{-4}$ corrected the problem.
 
\begin{table*}[t]
\caption{Sets of modes used to analyse the impact of the number and type of frequencies on the inversion results}
\label{tabmodes}
  \centering
\begin{tabular}{r | c | c | c | c | c | c | c | c}
\hline \hline
 & Set$_{1}$ & Set$_{2}$ & Set$_{3}$ & Set$_{4}$ & Set$_{5}$ & $16$CygA & $16$CygB \\ \hline
 $\ell=0$ & $n=9-28$ & $n=5-27$ & $n=9-34$ & $n=11-24$ & $n=11-26$ & $n=12-27$ & $n=13-26$ \\
 $\ell=1$ & $n=9-28$ & $n=5-27$ & $n=9-34$ & $n=11-24$ & $n=11-26$ & $n=11-27$ & $n=13-26$ \\
 $\ell=2$ & $n=9-28$ & $n=5-27$ & $n=9-34$ & $n=11-24$ & $n=11-26$ & $n=11-24$  & $n=12-25$ \\ 
 $\ell=3$ & $-$ & $-$ & $-$ & $n=9-20$ & $n=12-22$ & $n=15-21$ & $n=17-24$ \\ 
\hline
\end{tabular}
\end{table*}

\begin{table*}[t]
\caption{Inversion results for Target$_{3}$ using the $(u, \Gamma_{1})$ kernels and sets $1-5$ of Table \ref{tabmodes}.}
\label{tabresultsuGam3modes}
  \centering
\begin{tabular}{r | c | c | c | c | c | c }
\hline \hline
 & $\frac{t_{u}^{\mathrm{ref}}}{R_{ref}^{6}}$ $(g^{2}/cm^{6})$ & $\frac{t_{u}^{\mathrm{inv}}}{R_{ref}^ {6}}$ $(g^{2}/cm^{6})$ $(u,\Gamma_{1})$ & $\frac{t_{u}^ {\mathrm{obs}}}{R_{tar}^ {6}}$ $(g^{2}/cm^{6})$& $\varepsilon_{\mathrm{Avg}}^{u,\Gamma_{1}}$& $\varepsilon_{\mathrm{Cross}}^{u,\Gamma_{1}}$&$\varepsilon_{\mathrm{Res}}^{u,\Gamma_{1}}$\\ \hline
 $\mathrm{Set}_{1}$ & $5.855$ &$5.700\pm 0.161$&$5.538$& $2.805 \times 10^{-2}$& $-4.246 \times 10^{-5}$ & $4.881 \times 10^{-4}$\\ 
 $\mathrm{Set}_{2}$ &$5.888$ &$5.566\pm 0.088$&$5.538$& $4.062 \times 10^{-3}$& $1.869 \times 10^{-4}$& $3.505 \times 10^{-4}$ \\ 
  $\mathrm{Set}_{3}$ & $5.895$&$5.690\pm 0.146$&$5.538$&$2.79 \times 10^{-2}$& $9.25 \times 10^{-6}$&$6.8 \times 10^{-4}$ \\
 $\mathrm{Set}_{4}$ & $5.886$ &$5.570\pm 0.110$&$5.538$& $6.074\times 10^{-3}$& $2.859\times 10^{-4}$& $-1.893 \times 10^{-3}$\\ 
 $\mathrm{Set}_{5}$ & $5.968$ &$5.630\pm 0.105$&$5.538$& $1.644 \times 10^{-2}$& $-4.072\times 10^{-4}$& $4.714\times 10^{-4}$\\
\hline
\end{tabular}
\end{table*}
\begin{table*}[h]
\caption{Inversion results for the $16$CygA and $16$CygB clones using the $(u, \Gamma_{1})$ kernels.}
\label{tabresultsuGamCyg}
  \centering
\begin{tabular}{r | c | c | c | c | c | c }
\hline \hline
 & $\frac{t_{u}^{\mathrm{ref}}}{R_{ref}^{6}}$ $(g^{2}/cm^{6})$ & $\frac{t_{u}^{\mathrm{inv}}}{R_{ref}^ {6}}$ $(g^{2}/cm^{6})$ $(u,\Gamma_{1})$ & $\frac{t_{u}^ {\mathrm{obs}}}{R_{tar}^ {6}}$ $(g^{2}/cm^{6})$& $\varepsilon_{\mathrm{Avg}}^{u,\Gamma_{1}}$& $\varepsilon_{\mathrm{Cross}}^{u,\Gamma_{1}}$&$\varepsilon_{\mathrm{Res}}^{u,\Gamma_{1}}$\\ \hline
 $16$CygA (Full, $\theta=10^{-5}$) & $2.965$ & $2.891\pm 0.083$ & $2.885$ & $2.641\times 10^{-3}$& $-1.780\times 10^{-3}$& $1.442\times 10^{-3}$\\
  $16$CygA (Full, $\theta=10^{-4}$) & $2.965$ & $2.872\pm 0.036$ & $2.885$ & $-4.033\times 10^{-3}$& $-9.487\times 10^{-4}$& $7.149\times 10^{-4}$\\
 $16$CygA (Small, $\theta=10^{-5}$) & $2.965$ &$2.971\pm 0.083$&$2.885$& $3.117 \times 10^{-2}$& $-2.577 \times 10^{-3}$ & $5.000 \times 10^{-4}$\\ 
  $16$CygA (Small, $\theta=10^{-4}$) & $2.965$ &$2.906\pm 0.031$&$2.885$& $8.240 \times 10^{-3}$& $-1.778 \times 10^{-3}$ & $7.000 \times 10^{-4}$\\  
 $16$CygB (Full) &$4.540$ &$4.007\pm 0.095$&$3.783$& $4.547 \times 10^{-2}$& $-2.407 \times 10^{-4}$& $2.277 \times 10^{-2}$ \\ 
 $16$CygB (Small) & $4.540$&$4.295\pm 0.113$&$3.783$&$1.093 \times 10^{-1}$& $-1.715 \times 10^{-3}$&$1.138 \times 10^{-2}$ \\
\hline
\end{tabular}
\end{table*}
\section{Impact of the quality of the forward modelling process on the inversion results}\label{secmodelstudy}
\begin{figure*}[t]
	\flushleft
		\includegraphics[width=18cm]{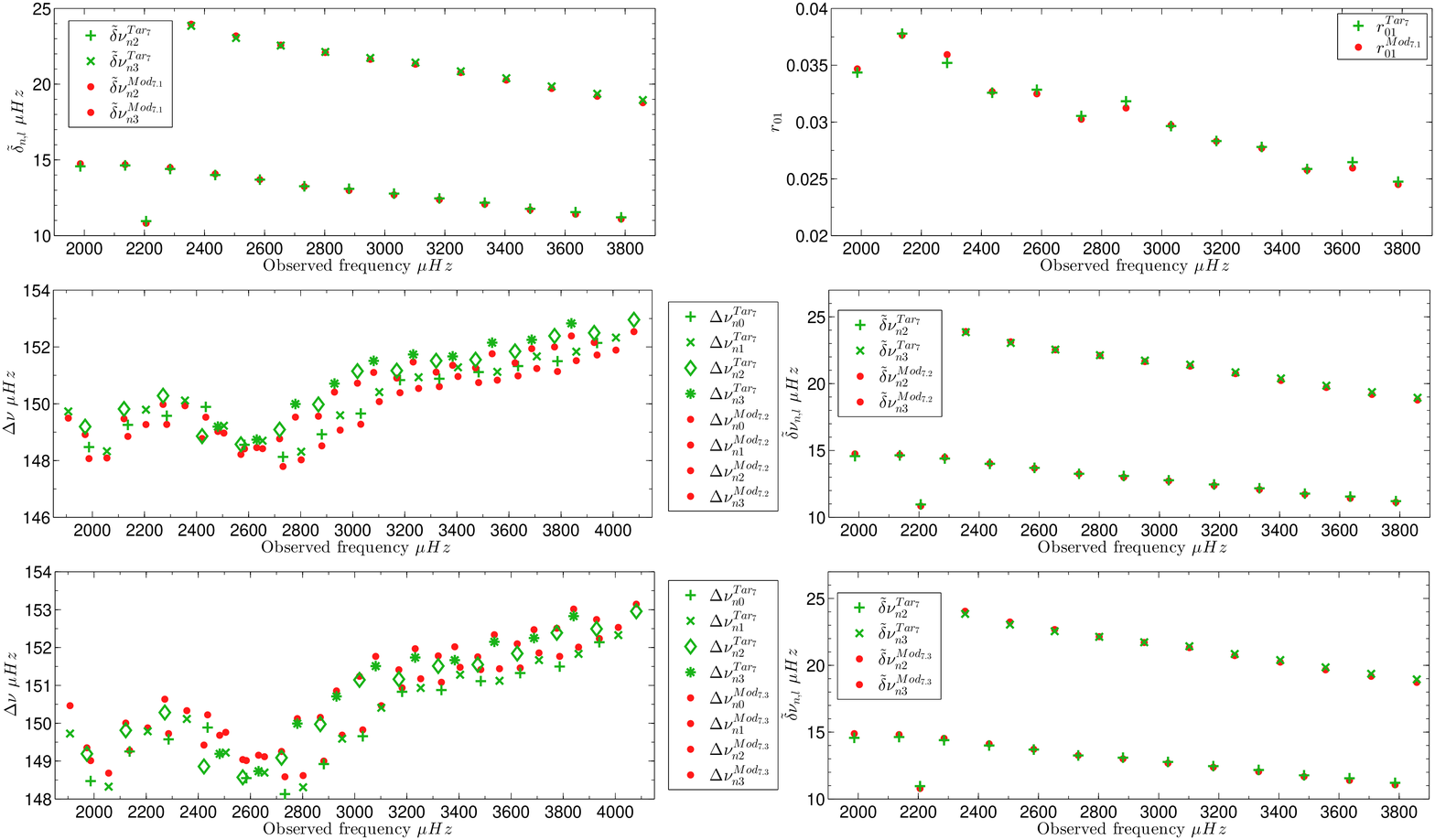}
	\caption{Results of the fit using the Levenberg-Marquardt algorithm for the first target as a function of the observed frequency: the upper panel is associated with $\mathrm{Model}_{7.1}$ which used the average large frequency separation and the individual $r_{01}$ and $\tilde{\delta} \nu_{nl}$ as constraints; the central panel is associated with $\mathrm{Model}_{7.2}$ which used individual $\Delta \nu_{n,l}$ and $\tilde{\delta} \nu_{n,l}$ as constraints; the lower panel is associated with $\mathrm{Model}_{7.3}$ which also used individual $\Delta \nu_{n,l}$ and $\tilde{\delta} \nu_{n,l}$ as constraints.}
		\label{figforwardtarget1}
\end{figure*}

In this section we present various inversion results using different criteria to select the reference model for the inversion. The previous results, only using the average large and small frequency separations as constraints for the mass and age of the model, are indeed a crude representation of the real capabilities of seismic modelling. It is well known that other individual frequency combinations can be used to obtain independent information on the core mixing processes and that we should adjust more than two parameters to describe the physical processes in stellar interiors.
\begin{table*}[t]
\caption{Characteristics of Target$_{7}$, Target$_{8}$ and of the models obtained with the Levenberg-Marquardt algorithm.}
\label{tabcaractargfor}
  \centering
\begin{tabular}{r | c | c | c | c | c | c | c}
\hline \hline
 & \textbf{Target$_{7}$}& \textbf{Model$_{7.1}$}& \textbf{Model$_{7.2}$}& \textbf{Model$_{7.3}$}& \textbf{Target$_{8}$}& \textbf{Model$_{8.1}$}&\textbf{Model$_{8.2}$}\\ \hline
 \textit{Mass ($\mathrm{M_{\odot}}$)}& $0.9$ &$0.933$& $0.908$ & $0.957$ & $1.0$ & $1.009$ & $1.029$\\
\textit{Radius ($\mathrm{R_{\odot}}$)}& $0.908$ & $0.919$& $0.912$ & $0.926$  &$1.17$& $1.18$ & $1.19$\\ 
\textit{Age ($\mathrm{Gyr}$)} &$3.075$&$3.34$ & $3.69$ & $3.45$ & $4.168$ & $4.322$& $4.489$\\ 
\textit{$T_{\mathrm{eff}}$ ($\mathrm{K}$)} & $5659$ & $5701$ & $5488$ & $5713$ &$6003$ & $5985$ &$5966$\\ 
\textit{$Z$} & $0.0122$&$0.0122$ & $0.0122$ & $0.0105$ & $0.0122$ & $0.0185$ & $0.0181$\\
\textit{$Y_{0}$} &$0.308$&$0.274$ & $0.269$ & $0.243$ & $0.3078$ & $0.323$&$0.305$\\
\textit{Abundances} &AGSS09&AGSS09&AGSS09&AGSS09&AGSS09&AGSS09 &AGSS09\\
\textit{$\alpha_{\mathrm{MLT}}$} & $1.522$ & $1.522$ & $1.297$ & $1.522$ & $1.522$ & $1.522$& $1.522$ \\
\textit{Diffusion coefficient} & $1.6$ &$0$ & $0$ & $0$ &$1.6$&$0$& $0$\\
\hline
\end{tabular}
\end{table*}
\begin{table*}[t]
\caption{Constraints and free parameters used for the Levenberg-Marquardt fit.}
\label{tabparamconstraints}
  \centering
\begin{tabular}{r | c | c }
\hline \hline
 &  \textbf{\textit{Constraints}} & \textbf{\textit{Parameters}}\\ \hline
 Model$_{7.0}$ & $< \Delta \nu_{n,l}(\nu) >$ + $< \tilde{\delta} \nu_{n,l} (\nu) >$ & Mass + Age \\
Model$_{7.1}$ & $\Delta \nu_{n,l}(\nu)$ + $\tilde{\delta} \nu_{n,l} (\nu)$ + $r_{01}(\nu)$& Mass + Age + $Y_{0}$ \\
Model$_{7.2}$ &$\Delta \nu_{n,l}(\nu)$ + $\tilde{\delta} \nu_{n,l}(\nu)$ & Mass + Age + $Y_{0}$+ $\alpha_{\mathrm{MLT}}$\\ 
Model$_{7.3}$ &$\Delta \nu_{n,l}(\nu)$ + $\tilde{\delta} \nu_{n,l}(\nu)$&Mass + Age + $Y_{0}$ + $Z_{0}$\\ 
Model$_{8.1}$ & $<\Delta \nu>$ + $r_{02}(\nu)$ & Mass + Age + $Y_{0}$ + $Z_{0}$\\ 
Model$_{8.2}$ & $<\Delta \nu>$ + $r_{02}(\nu)$ + $r_{01}(\nu)$&Mass + Age + $Y_{0}$ + $Z_{0}$\\
\hline
\end{tabular}
\end{table*}
To carry out these test cases, we build two target models including microscopic diffusion. As previously done, we did not include this process in the reference models obtained by fitting the $56$ ``observed'' frequencies. We used the modes $l=0, n=12-25$; $l=1, n=11-25$; $l=2,n=11-26$; $l=3, n=14-24$. The characteristics of the targets are summarized in Table \ref{tabcaractargfor} along with those of the best models obtained through seismic modelling. Table \ref{tabparamconstraints} contains information on the various constraints and free parameters used for the fit. We used various seismic constraints such as the individual large and small frequency separations, the individual $r_{01}$ and $r_{02}$, defined as:
\begin{align}
r_{02}&=\frac{\nu_{n-1,2}-\nu_{n,0}}{\nu_{n,1}-\nu_{n-1,1}}, \\
r_{01}&=\frac{\nu_{n,0}-2\nu_{n,1}+\nu_{n+1,0}}{2(\nu_{n+1,0}-\nu_{n,0})}.
\end{align}
The free parameters were chosen to match what is done when trying to fit observations, although the final quality of the fit is much higher than what one expects from an observed case, as is illutrasted in Fig. \ref{figforwardtarget1} for Target$_{7}$\footnote{We did not present the fit of the individual large frequency separations for Model$_{7.1}$ to avoid redundancy with Model$_{7.2}$ and Model$_{7.3}$.} and Tab. \ref{tabforwardtarget2} for Target$_{8}$, where one should note that the arithmetic average of the large frequency separations were fitted to within $1\%$ in addition to the individual quantities plotted in Tab. \ref{tabforwardtarget2}. In all these cases the inversion improved the value of $t_{u}$. In some cases, the acoustic radius was not well fitted by the forward modelling process but the inversion could improve its determination. 
\begin{figure*}[t]
	\flushleft
		\includegraphics[width=18cm]{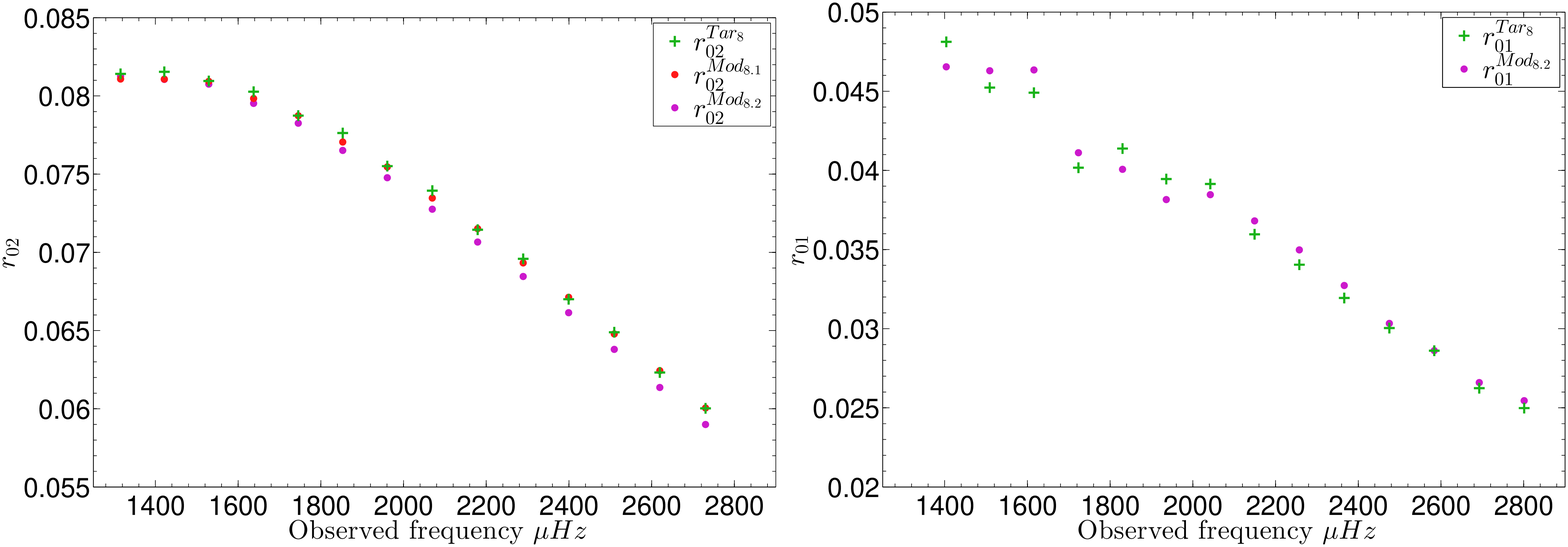}
	\caption{Results of the fit using the Levenberg-Marquardt algorithm for the second target as a function of the observed frequency (colour online): the red dots are associated with $\mathrm{Model}_{8.1}$ which used individual $r_{02}$ as constraints and the purple dots are associated with $\mathrm{Model}_{8.2}$ which used individual $r_{02}$ and $r_{01}$ as constraints.}
		\label{figforwardtarget2}
\end{figure*} 
Table \ref{tabresforward2} illustrates the results of combined inversions whereas Table \ref{tabresultsforward} gives the results for all test cases. One could point out that the error contributions are not that different from what was obtained using only the average large and small frequency separations as constraints. However, our previous analysis of the dependence in the degree and radial order of the modes has shown us that having low $n$ and $l=3$ modes was the best way to ensure accuracy. In that sense, the frequency set used for these test cases is of lower quality and but this has been compensated for by the forward modelling process. To further illustrate the importance of the model selection, we repeat in Table \ref{tabresultsforward} the results obtained by simply ajusting $ <\Delta \nu >$ and $< \tilde{\delta} \nu >$ that we denote as Model$_{7.0}$. We see from the results for Target$_{7}$ that the dominant source of error contribution, $\varepsilon_{\mathrm{Avg}}$ is $\pm 6$ to $10$ times smaller than what is for the Model$_{7.0}$. It is thus clear that using the information given by individual frequencies is crucial to ensure accurate results in observed cases.
\\
\\
The necessity of an acoustic radius inversion results from two aspects in the selection of the reference models. The first one is present in $\mathrm{Model}_{7.2}$; the change of $\alpha$ induced during the fit had an important impact on the upper regions and therefore a change in the acoustic radius was observable. The second one is present in $\mathrm{Model}_{8.1}$ and $\mathrm{Model}_{8.2}$, where the observational constraints were sensitive to core regions, except for the arithmetic average of the large frequency separations. In this case, the upper regions are thus less constrained and the inversion is still necessary. However, we note that in most cases the acoustic radius of the reference model was very accurate. This is due to the lack of surface effects in the target models. If we were to include non-adiabatic computations or differences in the convection treatment, differences would be seen in the acoustic radii of the targets and reference models but it is clear from these test cases that the acoustic radius combined with $t_{u}$ alone is not sufficient to disentangle between effects of differences in helium abundance and effects of microscopic diffusion. We also note that when using the individual $r_{02}$ along with the $< \Delta \nu>$ (the test case of Model$_{8.1}$), we obtain a very good fit of $t_{u}$ with the reference model. This is a consequence of the fact that $r_{02}$ is very sensitive to core regions and therefore the core characteristics are well reproduced. However, with the acoustic radius inversion, we note that the surface regions are not well fitted, and if we include $r_{01}$, as is done in the test case of Model$_{8.2}$, we obtain a better fit of the acoustic radius, but $t_{u}$ is then less accurate. The inversion of the indicator $t_{u}$ informs us that the core regions are not well reproduced in this reference model. We can see that Model$_{8.1}$ or Model$_{7.2}$ and Model$_{7.3}$ reproduce better the indicator $t_{u}$. However, in all these cases the acoustic radius was not properly reproduced. Therefore the combined inversions indicate that something is wrong with the set of free parameters used because we cannot fit properly surface and core regions simultaneously, although the fit of the seismic constraints with the Levenberg-Marquardt algorithm is excellent in all these cases.
\\
\\
We also mention that in all test cases carried out here, we did not consider the first age indicator $t$ from \citet{Buldgen}. In fact, the indicator $t$ could also provide accurate results for Target$_{7}$. However, its inaccuracy for older models has been observed during this entire study and we recommend to limit its use to young stars\footnote{We chose not to present these results to focus our study on the $t_{u}$ indicator.}, where it can provide valuable information provided the kernels are well optimised.
\begin{table*}[t]
\caption{Inversion results for various fits with the Levenberg-Marquardt algorithm.}
\label{tabresultsforward}
  \centering
\begin{tabular}{r | c | c | c | c | c | c }
\hline \hline
 & $\frac{t_{u}^{\mathrm{ref}}}{R_{ref}^{6}}$ $(g^{2}/cm^{6})$ & $\frac{t_{u}^{\mathrm{inv}}}{R_{ref}^ {6}}$ $(g^{2}/cm^{6})$ $(u,\Gamma_{1})$ & $\frac{t_{u}^ {\mathrm{obs}}}{R_{tar}^ {6}}$ $(g^{2}/cm^{6})$& $\varepsilon_{\mathrm{Avg}}^{u,\Gamma_{1}}$& $\varepsilon_{\mathrm{Cross}}^{u,\Gamma_{1}}$&$\varepsilon_{\mathrm{Res}}^{u,\Gamma_{1}}$\\ \hline
   Model$_{7.0}$ & $7.626$ & $7.000\pm 0.24$ & $6.703$ & $4.952\times 10^{-2}$& $2.94\times 10^{-4}$& $-3.389\times 10^{-3}$\\
 Model$_{7.1}$ & $6.562$ & $6.657\pm 0.122$ & $6.703$ & $-1.4\times 10^{-3}$& $-2.407\times 10^{-4}$& $-2.777\times 10^{-4}$\\
 Model$_{7.2}$ & $6.393$ & $6.669\pm 0.123$ & $6.703$ & $-6.597\times 10^{-3}$& $-1.292\times 10^{-4}$& $-2.478\times 10^{-5}$\\
 Model$_{7.3}$ & $6.467$ &$6.667\pm 0.121$&$6.703$& $-2.571 \times 10^{-3}$& $-3.322 \times 10^{-4}$ & $-5.661 \times 10^{-4}$\\ 
 Model$_{8.1}$ & $3.450$ &$3.651\pm 0.092$&$3.568$& $2.458 \times 10^{-2}$& $-2.302 \times 10^{-4}$ & $-1.745 \times 10^{-3}$\\  
 Model$_{8.2}$ &$3.337$ &$3.637\pm 0.096$&$3.568$& $2.089 \times 10^{-2}$& $-6.553 \times 10^{-4}$& $-1.568 \times 10^{-3}$ \\ 
\hline
\end{tabular}
\end{table*}

\begin{table*}[t]
\caption{Combined $(\tau,t_{u})$ inversion results for various fits with the Levenberg-Marquardt algorithm.}
\label{tabresforward2}
  \centering
\begin{tabular}{r | c | c | c | c | c | c }
\hline \hline
 & $\frac{t_{u}^{\mathrm{ref}}}{R_{ref}^{6}}$ $(g^{2}/cm^{6})$ & $\frac{t_{u}^{\mathrm{inv}}}{R_{ref}^ {6}}$ $(g^{2}/cm^{6})$ $(u,\Gamma_{1})$ & $\frac{t_{u}^ {\mathrm{obs}}}{R_{tar}^ {6}}$ $(g^{2}/cm^{6})$& $\tau_{\mathrm{ref}} (s)$& $\tau_{\mathrm{inv}} (s)$&$\tau_{\mathrm{obs}} (s)$\\ \hline
   Model$_{7.2}$ & $6.393$ & $6.657\pm 0.092$ & $6.703$ & $3230$& $3223 \pm 0.028$& $3222$\\
 Model$_{8.1}$ & $3.450$ & $3.651\pm 0.096$ & $3.568$ & $4509$& $4450 \pm 0.028$& $4442$\\
 Model$_{8.2}$ & $3.337$ & $3.637\pm 0.120$ & $3.568$ & $4517$& $4448 \pm 0.017$& $4442$\\
\hline
\end{tabular}
\end{table*}
\section{Conclusion} \label{secconclusion}
In this article, we have presented a new approach to constrain mixing processes in stellar cores using the SOLA inversion technique. We used the framework presented in \citet{Buldgen} to develop an integrated quantity, denoted $t_{u}$, sensitive to the effects of stellar evolution and to the impact of additional mixing processes or mismatches in the chemical composition of the core. We based our choice solely on structural effects and considerations about the variational principle and the ability of the kernels to fit their targets. 
\\
\\
The derivation of this new quantity was made possible by the use of the approach of \citet{Masters} to derive new structural kernels in the context of asteroseismology by solving an ordinary differential equation. We discussed the problem of the intrinsic scaling effect presented in \citet{Basusca} and discussed how it could affect the indicator $t_{u}$. We tested its sensitivity to various physical changes between the target and the reference model and demonstrated that SOLA inversions are able to significantly improve the accuracy with which $t_{u}$ is determined, thereby indicating whether there is a problem in the core regions of the reference model.
\\
\\
We also analysed the importance of the number and type of modes in the observational data and concluded that the accuracy of an inversion of the indicator $t_{u}$ increased with multiple values of the degree, $\ell$, and low values of the radial order, $n$. In that sense, we emphasize that the observation of $\ell=3$ modes is important for the inversion of the indicator $t_{u}$ since it can improve the accuracy without the need of low $n$ modes. Such modes are difficult to observe. Indeed, only a few octupole modes have been detected for around $15 \%$ of solar-like stars with Kepler. The use of other observational facilities, such as the SONG network \textbf{\citep{SONG}}, might help us obtain richer oscillation spectra as far as the octupole modes are concerned. The test cases for the $16$CygA and $16$CygB clones demonstrated that our method was applicable to the current observational data and one could still carry out an inversion of $t_{u}$ without these modes. However, it is clear that this method will only be applicable to the best observational cases with Kepler, Plato or SONG.
\\
\\
We also analysed the impact of the selection of the reference model on the inversion results and concluded that using individual frequency combinations is far more efficient in terms of accuracy and stability of the inversion results. However, we also noticed in supplementary tests that there was what could be called a ``resolution limit'' for the $t_{u}$ inversion which depends on the magnitude of the differences in the physics between the reference model and its targets but also on the weight given to the core in the selection of the reference model (see for example the case of Model$_{8.1}$ and the associated discussion). This leads to the conclusion that supplementary independent integrated quantities should be derived to help us disentangle between various physical effects and improve our sensitivity to the physics of stellar interiors. Nevertheless, the test cases of Sect. \ref{secmodelstudy} showed that the SOLA method was much more sensitive than the forward modelling process used to select the reference model (here a Levenberg-Marquardt algorithm) and could indicate whether the set of free parameters used to describe the model is adequate. 
\begin{acknowledgements}
G.B. is supported by the FNRS (``Fonds National de la Recherche Scientifique'') through a FRIA (``Fonds pour la Formation à la Recherche dans l'Industrie et l'Agriculture'') doctoral fellowship. D.R.R. is currently funded by the European Community's Seventh Framework Programme (FP7/2007-2013) under grant
agreement no. 312844 (SPACEINN), which is gratefully acknowledged. This article made use of an adapted version of InversionKit, a software developed in the context of the HELAS and SPACEINN networks, funded by the European Commissions's Sixth and Seventh Framework Programmes.
\end{acknowledgements}
\bibliography{biblioarticle2}
\end{document}